\documentclass[onecolumn,11pt]{article}
\pdfoutput=1
\usepackage[top=1in, bottom=1in, left=1in, right=1in]{geometry}
\usepackage{amsmath,amsfonts,amscd,amssymb}
\usepackage{graphicx}
\usepackage{subfigure}
\usepackage{epstopdf}
\usepackage{overpic}
\usepackage{arcs}
\usepackage{color}
\usepackage{setspace}
\usepackage{palatino} 
\usepackage{relsize}
\usepackage{lmodern}
\usepackage{slantsc}
\usepackage{float}
\usepackage{hyperref}
\usepackage{comment}
\usepackage[numbers,sort&compress]{natbib}

\definecolor{offlinered}{RGB}{115, 0, 0}
\definecolor{onlineblue}{RGB}{4, 79, 149}

\DeclareSymbolFont{yhlargesymbols}{OMX}{yhex}{m}{n}
\DeclareMathAccent{\wideparen}{\mathord}{yhlargesymbols}{"F3}

\newenvironment{mat}{\left[\begin{array}{ccccccccccccccc}}{\end{array}\right]}
\newcommand\bcm{\begin{mat}}
\newcommand\ecm{\end{mat}}

\newenvironment{rmat}{\left[\begin{array}{rrrrrrrrrrrrr}}{\end{array}\right]}
\newcommand\brm{\begin{rmat}}
\newcommand\erm{\end{rmat}}

\newcommand{\eql}{\begin{equation}\label}

\begin{document}
\title{\vspace{-.1in}\huge{Stable \textit{a posteriori} LES of 2D turbulence using convolutional neural networks: Backscattering analysis and generalization to higher $Re$ via transfer learning}}

\author{\normalsize{Yifei Guan$^1$\thanks{yifei.guan@rice.edu}, Ashesh Chattopadhyay$^1$, Adam Subel$^1$, and Pedram Hassanzadeh$^{1,2}$\thanks{pedram@rice.edu}}\\
\footnotesize{$^1$Department of Mechanical Engineering, Rice University, Houston, TX, 77005, United States}\\
\footnotesize{$^2$Department of Earth, Environmental and Planetary Sciences, Rice University, Houston, TX, 77005, United States}\\
}
\date{}
\maketitle

\maketitle

\begin{abstract}
There is a growing interest in developing data-driven subgrid-scale (SGS) models for large-eddy simulation (LES) using machine learning (ML). In \textit{a priori} (offline) tests, some recent studies have found ML-based data-driven SGS models that are trained on high-fidelity data (e.g., from direct numerical simulation, DNS) to outperform baseline physics-based models and accurately capture the inter-scale transfers, both forward (diffusion) and backscatter. While promising, instabilities in \textit{a posteriori} (online) tests and inabilities to generalize to a different flow (e.g., with a higher Reynolds number, $Re$) remain as major obstacles in broadening the applications of such data-driven SGS models. For example, many of the same aforementioned studies have found instabilities that required often ad-hoc remedies to stabilize the LES at the expense of reducing accuracy. Here, using 2D decaying turbulence as the testbed, we show that deep fully convolutional neural networks (CNNs) can accurately predict the SGS forcing terms and the inter-scale transfers in \textit{a priori} tests, and if trained with enough samples, lead to stable and accurate \textit{a posteriori} LES-CNN. Further analysis attributes these instabilities to the disproportionally lower accuracy of the CNNs in capturing backscattering when the training set is small. We also show that transfer learning, which involves re-training the CNN with a small amount of data (e.g., $1\%$) from the new flow, enables accurate and stable \textit{a posteriori} LES-CNN for flows with $16\times$ higher $Re$ (as well as higher grid resolution if needed). These results show the promise of CNNs with transfer learning to provide stable, accurate, and generalizable LES for practical use.
\end{abstract}

\section{Introduction}
Accurate simulations of turbulent flows are of critical importance for predicting and understanding various engineering and natural systems. However, the direct numerical simulation (DNS) of the Navier-Stokes equations remains computationally prohibitive for many real-world applications because DNS requires resolving (i.e., directly solving for) all the relevant spatial and temporal scales. These scales might span several orders of magnitude, e.g., from the domain length down to the Kolmogorov scale~\cite{moin1998direct,pope2001turbulent}. Large-eddy simulation (LES) offers a balance between accuracy and computational cost, since in LES, only the part of the inertial range containing the large-scale structures is resolved on a coarse-resolution grid and the effects of the subgrid-scale (SGS) eddies are parameterized, in terms of the resolved flow, using a SGS model~\cite{pope2001turbulent,sagaut2013multiscale}. As a result, the quality of the solutions from LES highly depends on the quality of the SGS model. Consequently, formulating accurate SGS models for LES has been an active area of research for the past few decades in different disciplines~\cite[e.g.,][]{meneveau2000scale,sagaut2013multiscale,dipankar2015large,sarlak2015role,pressel2017numerics,schneider2017earth}. Below, we briefly describe some of the key physics-based SGS models and their major shortcomings, which have motivated the recent interest in using machine learning (ML) to find data-driven SGS models. Then, we discuss some of the advances in data-driven SGS modeling as well as the main challenges, some of which we aim to address in this paper. \\

In his pioneering work on developing one of the first global climate models, Smagorinsky proposed a physics-based SGS model for LES in 1963~\cite{smagorinsky1963general}. In this model (SMAG, hereafter), effects of the SGS eddies are parameterized as a function of the resolved flow using a scale-selective dissipative model that consists of a positive eddy viscosity $\nu_e$ and second-order diffusion. Since then, the SMAG model and its variants have been widely used in different disciplines, for example, to simulate weather and climate variability, combustion, multiphase flows, wind farms, and magnetohydrodynamics~\cite[e.g.,][]{arakawa1977computational,piomelli1999large,knaepen2004large,sagaut2006large,pitsch2006large,fox2008can,fox2012large,tan2017large,stevens2018comparison}. Such purely diffusive SGS models lead to numerically stable LES; however, they might not correctly capture the inter-scale physical processes such as energy (and enstrophy) transfers. These models often include second-order dissipation but higher orders can also contribute to the forward transfer, i.e., transfer from the resolved scales to the subgrid scales~\cite[]{meneveau2000scale}. Furthermore, while in the mean, the transfer is forward and the role of SGS processes is indeed dissipative, it is known that locally there can be transfer from the subgrid scales to the resolved scales. This process is referred to as backscattering, which is missing from purely diffusive SGS models~\cite{pope2001turbulent}. \\

Backscattering has been found to play a significant role in various fluid flows, and extensive work has been done in different disciplines to account for it in physics-based SGS models~\cite[e.g.,][]{leslie1979application,zhou1991eddy,mason1992stochastic,carati1995representation,domaradzki1987analysis,kerr1996small,thuburn2014cascades,o2014subgrid,khani2016backscatter,shinde2020proper}. For example, Piomelli~\textit{et al.}~\cite{,piomelli1990large,piomelli1991subgrid} showed that the lack of energy backscattering in LES could lead to inaccurate prediction of the perturbation growth in transitional wall-bounded flows. Backscattering has been also found to be important in geophysical turbulence, which has implications for modeling atmospheric and oceanic circulations and weather/climate predictions~\cite{shutts2005kinetic,berner2008impact,nadiga2010stochastic,grooms2015numerical,jansen2015energy,rasp2018deep,hewitt2020resolving}. To improve the SMAG model and account for backscattering, Germano~\textit{et al.}~\cite{germano1991dynamic} developed a dynamic approach to compute the eddy viscosity, which could lead to $\nu_e<0$ (anti-diffusion) and account for backscattering. While this model (known as dynamic Smagorinsky; DSMAG hereafter) and its variants were shown to accurately represent many aspects of inter-scale energy transfers, it could also lead to numerical instabilities~\cite{lilly1992proposed,meneveau1997dynamic}. As a result, later modifications were proposed to enforce $\nu_e \ge 0$ as a tradeoff between numerical stability and backscattering~\cite{zang1993dynamic}. Adding stochasticity to eddy-viscosity SGS models as well as other approaches have been proposed to improve their accuracy (e.g., account for backscattering) while maintaining stability~\cite[e.g.,][]{domaradzki1987analysis,chasnov1991simulation,carati1995representation,domaradzki1997subgrid,liu2011modification,mana2014toward,jansen2014parameterizing,jansen2015energy}. Despite these efforts, the need for better SGS models that accurately account for both forward and backscatter transfers remains. As a motivating example, the parameterizations currently used in global climate models do not account for kinetic energy backscattering~\cite{hewitt2020resolving}. \\

In the past few years, there has been a rapidly growing interest in using ML methods to improve the modeling and analysis of chaotic systems and turbulent flows~\cite[e.g.,][]{pathak2018model,wan2018data,raissi2019physics,bar2019learning,wu2020enforcing,mohan2020spatio,chattopadhyay2020ESN,chattopadhyay2020analog,frezat2020physical,maulik2020probabilistic,sirignano2020dpm,pathak2020using,novati2021automating,Kashinath2021}; also see the recent review papers on this topic~\cite{duraisamy2019turbulence,brunton2020machine,pandey2020perspective,beck2020perspective,duraisamy2020machine}. Specific to SGS modeling (for LES or other approaches), a number of studies have aimed to obtain better estimates for the parameter(s) of physics-based SGS models, such as $\nu_e$, from high-fidelity data (e.g., DNS or observations)~\cite[][]{sarghini2003neural,schneider2017earth,wu2018physics,dunbar2020calibration,maulik2020turbulent,souza2020uncertainty}. Alternatively, a growing number of recent papers have aimed to learn a data-driven SGS model from high-fidelity data, often in a non-parametric fashion, i.e., without any prior assumption about the model's structural/functional form~\cite[e.g.,][]{ling2016reynolds,parish2016paradigm,gamahara2017searching,wang2017physics,wang2018investigations,salehipour2019deep,aandorp2020data,taghizadeh2020turbulence,frezat2020physical,pawar2020interface,portwood2021interpreting}. In the studies from the latter category that focused on LES, a variety of canonical fluid systems and different approaches (e.g., local vs. non-local) have been investigated. In the local approach, which often employs multilayer perceptron artificial neural networks (ANNs), the SGS term (stress tensor or its divergence) at a grid point is estimated in terms of the resolved flow at or around the same grid point. For example, Maulik~\textit{et al.}~\cite{maulik2019subgrid} and Xie~\textit{et al.}~\cite{xie2019artificial,xie2019artificial2} have, respectively, studied 2D decaying homogenous isotropic turbulence (2D-DHIT) and 3D incompressible and compressible turbulence using this approach (also, see \cite{xie2021artificial}). In the non-local approach, which often employs variants of convolutional neural networks (CNNs), the SGS term over the entire domain is estimated in terms of the resolved flow in the entire domain to account for potential spatial correlations (e.g., due to coherent structures) and non-homogeneities in the system. For example, Zanna and Bolton~\cite{bolton2019applications,zanna2020data}, Beck and colleagues~\cite{beck2019deep,kurz2020machine}, Pawar~\textit{et al.}~\cite{pawar2020priori}, and Subel~\textit{et al.}~\cite{subel2020data}  have used this approach for ocean circulation, 3D-DHIT, 2D-DHIT, and forced 1D Burgers' turbulence, respectively.  \\

In \textit{a priori} (offline) tests, in which the accuracy of the SGS model in estimating the SGS term as a function of the resolved flow is evaluated, some of these studies have found the data-driven SGS models to accurately account for inter-scale transfers (including backscattering) and outperform physics-based models such as SMAG and DSMAG~\cite{bolton2019applications,maulik2019subgrid,zhou2019subgrid,zanna2020data,pawar2020priori}. However, most of the same studies have also found that in \textit{a posteriori} (online) tests, in which the data-driven SGS model is coupled with a coarse-resolution numerical solver, the LES model is unstable, leading to numerical blow-up or physically unrealistic flows~\cite{maulik2019subgrid,beck2019deep,zhou2019subgrid,kurz2020machine,zanna2020data,stoffer2020development,beck2020perspective,xie2020modeling}. While the reason(s) for these instabilities remain unclear, a number of remedies have been proposed, e.g., post-processing of the trained SGS model to remove backscattering or to attenuate the SGS feedback into the numerical solver, or combining the data-driven model with an eddy viscosity model~\cite{maulik2019subgrid,beck2019deep,zhou2019subgrid,zanna2020data} (also, see the excellent review by Beck and Kurz~\cite{beck2020perspective}). However, such remedies include ad-hoc components and often substantially take away the advantages gained from the non-parametric, data-driven approach.  \\

Instabilities in \textit{a posteriori} tests remain a major challenge to broadening the applications of ML-based data-driven SGS models for LES. Another major challenge is the generalization capability of the data-driven SGS models beyond the flow from which the training data are obtained, e.g., extrapolation to turbulent flows with higher Reynolds numbers ($Re$). The ability to generalize is important for at least two reasons: i) High-fidelity data from usually expensive simulations (e.g., DNS) are needed to train data-driven SGS models and given the sharp increase in the computational cost of DNS with $Re$, the ability to effectively extrapolate to higher $Re$ makes data-driven SGS models much more useful in practice; and ii) Some level of generalization capability in the data-driven SGS models is essential for the LES models to be robust and trustworthy. However, it is known that such extrapolations are challenging for neural networks in general~\cite{krueger2020out}. In LES modeling, \textit{a priori} tests in 3D-DHIT have shown that the performance of data-driven SGS models degrades when applied to $Re$ higher than the one for which the model is trained. In \textit{a posteriori} tests with multi-scale Lorenz 96 systems~\cite{chattopadhyay2020data} and forced 1D Burgers' turbulence~\cite{subel2020data}, we found inaccurate generalization to more chaotic systems or flows with higher $Re$, particularly in terms of short-term prediction and re-producing the long-term statistics of rare events. However, in both studies, we also found that transfer learning, which involves re-training (part of) the already trained neural network using a small amount of data from the new system~\cite{yosinski2014transferable}, enables accurate generalization, e.g., to $10\times$ higher $Re$~\cite{chattopadhyay2020data,subel2020data}. While promising, the effectiveness of transfer learning in enabling generalization in more complex turbulent flows needs to be investigated.  \\

Building on these earlier studies, here we use a deep fully CNN architecture to build a non-local data-driven SGS model for a 2D-DHIT system using DNS data, and aim to
\begin{enumerate}
  \item Examine the accuracy of this SGS model in \textit{a priori} (offline) tests, with regard to both predicting the SGS terms and capturing inter-scale transfers,
  \item Evaluate the accuracy and stability of LES with this SGS model (LES-CNN) in \textit{a posteriori} (online) tests, both in terms of short-term predictions and long-term statistics,
  \item Assess the effectiveness of transfer learning in enabling accurate and stable generalization of LES-CNN to higher $Re$ (up to $16\times$). We also show generalization to higher grid resolutions by adding an encoder-decoder architecture to the CNN.
\end{enumerate}
For (a) and (b), we also present results from the SMAG and DSMAG models as well as a local ANN-based data-driven SGS model. \\

The remainder of this paper is structured as follows. Governing equations of the 2D-DHIT system, the filtered equations, and the DNS and LES numerical solvers are presented in Section~\ref{sec:eqs}, followed by descriptions of the data-driven SGS models (training data and the CNN and ANN architectures) and the physics-based SGS models (SMAG and DSMAG) in Section~\ref{LES methods}. Results of the \textit{a priori} and \textit{a posteriori} tests as well as generalization to higher $Re$ and/or resolutions via transfer learning are presented in Section~\ref{sec:results}. Conclusions and future work are discussed in Section~\ref{sec:conclusions}.

\section{DNS and LES: Governing equations and numerical solvers} \label{sec:eqs}
\subsection{Governing equations}
The dimensionless governing equations of 2D-DHIT in the vorticity ($\omega$) and streamfunction ($\psi$) formulation in a doubly periodic $x-y$ domain are
\begin{subequations}\label{Navier-Stokes}
\begin{eqnarray}
\frac{\partial \omega}{\partial t} + \mathcal{N}(\omega,\psi)&=&\frac{1}{Re}\nabla^2\omega\label{NS1}, \label{eq:NS}\\
\nabla^2\psi &=& -\omega\label{NS2}, \label{eq:omega}
\end{eqnarray}
\end{subequations}
where the nonlinear term $\mathcal{N}(\omega,\psi)$ represents advection
\begin{eqnarray}\label{NL}
\mathcal{N}(\omega,\psi)&=&\frac{\partial \psi}{\partial y}\frac{\partial \omega}{\partial x} - \frac{\partial \psi}{\partial x}\frac{\partial \omega}{\partial y}.
\end{eqnarray}
2D turbulence is a fitting prototype for many large-scale geophysical and environmental flows (where rotation and/or stratification dominate) and has been widely used as a testbed for novel techniques, including ML-based SGS modeling \cite[e.g.,][]{vallis2017atmospheric,tabeling2002two,chandler2013invariant,thuburn2014cascades,verkley2019maximum}. In DNS, as discussed in detail in Section~\ref{sec:num}, Eqs.~(\ref{eq:NS})-(\ref{eq:omega}) are numerically solved at high spatio-temporal resolutions. \\

To find the equations for LES, we apply filtering (denoted by $\overline{(\cdot)}$ and defined later) to Eqs.~\eqref{NS1}-\eqref{NS2} to obtain
\begin{subequations}\label{Fitlerd-Navier-Stokes}
\begin{eqnarray}
\frac{\partial \overline{\omega}}{\partial t} + \mathcal{N}(\overline{\omega},\overline{\psi})&=&\frac{1}{Re}\nabla^2\overline{\omega}+\underbrace{\mathcal{N}(\overline{\omega},\overline{\psi}) - \overline{\mathcal{N}({\omega},{\psi})}}_{\Pi}\label{FNS1},\\
\nabla^2\overline{\psi} &=& -\overline{\omega}\label{FNS2}.
\end{eqnarray}
\end{subequations}
Note that in deriving these equations, we assume that the filter commutes with the spatial (and temporal) derivative operators, which is the case for  commonly used filters such as box, sharp spectral, and Gaussian filters \cite{pope2001turbulent,sagaut2006large}; the latter is used in this work (see Section~\ref{sec:filtering}). As discussed in Section~\ref{sec:num}, ths numerical solution of Eqs.~\eqref{FNS1}-\eqref{FNS2} requires spatio-temporal resolutions lower than those of the DNS. However, the SGS forcing term, $\Pi$, includes the effects of the small-scale eddies that have been truncated due to filtering/coarse-graining and are not resolved in LES. As a result, $\Pi$ has to be estimated solely based on the resolved variables ($\overline{\omega},\overline{\psi}$) to close Eqs.~\eqref{FNS1}-\eqref{FNS2}, a problem that is at the heart of turbulence modeling \cite{pope2001turbulent}. \\

In most physics-based models, such as those using eddy viscosity, $\Pi$ is modeled as a purely diffusive process (SMAG and DSMAG are described in Section~\ref{sec:smag}). In data-driven approaches, such as the one pursued here and discussed in Sections~\ref{sec:cnn} and \ref{sec:ann}, the aim is to learn the relationship between ($\overline{\omega},\overline{\psi}$) and $\Pi$ in DNS data using methods such as deep neural networks, without any prior assumptions about the functional form of this relationship.

\subsection{Numerical solvers} \label{sec:num}
For DNS, we solve Eqs.~\eqref{NS1}-\eqref{NS2} in a doubly periodic square domain with $L\times L = [0,2\pi]\times[0,2\pi]$. A Fourier-Fourier pseudo-spectral solver is used along with second-order Adams-Bashforth and Crank-Nicolson time-integration schemes for the advection and viscous terms, respectively. The computational grid has uniform spacing $\Delta_\mathrm{DNS} = 2\pi/N_\mathrm{DNS}$, where $N_\mathrm{DNS}$ is the number of grid points in each direction. We use $N_\mathrm{DNS}=2048$ for $Re = 8000,32000, \text{and } 64000$, and $N_\mathrm{DNS}=3072$ for $Re = 128000$. The time-stepping size $\Delta t_\mathrm{DNS}= 10^{-4}$ ($\Delta t_\mathrm{DNS} = 5\times10^{-5}$) is used for $N_\mathrm{DNS}=2048$ ($N_\mathrm{DNS}=3072$). Following Refs.~\cite{maulik2017stable,maulik2019subgrid}, the initial condition of each DNS run is a random vorticity field but with the same prescribed energy spectrum (see Appendix~\ref{Initial conditions of DNS} for details). For each of the $Re$ mentioned above, we conducted $15$ independent DNS runs from random initial conditions.  \\

The numerical solver is implemented in Python using CUDA GPU computing. We use equal numbers of GPU blocks as the resolution in each direction such that only one GPU thread in each block is assigned for the computation on one computational grid point. The fast Fourier transform (FFT) and inverse fast Fourier transform (iFFT) operations are performed using the cuFFT library. Double-precision floating-point arithmetic is used for all numerical solvers. \\

\begin{figure}[t]
\vspace{.1in}
 \centering
 \begin{overpic}[width=1\linewidth,height=1\linewidth]{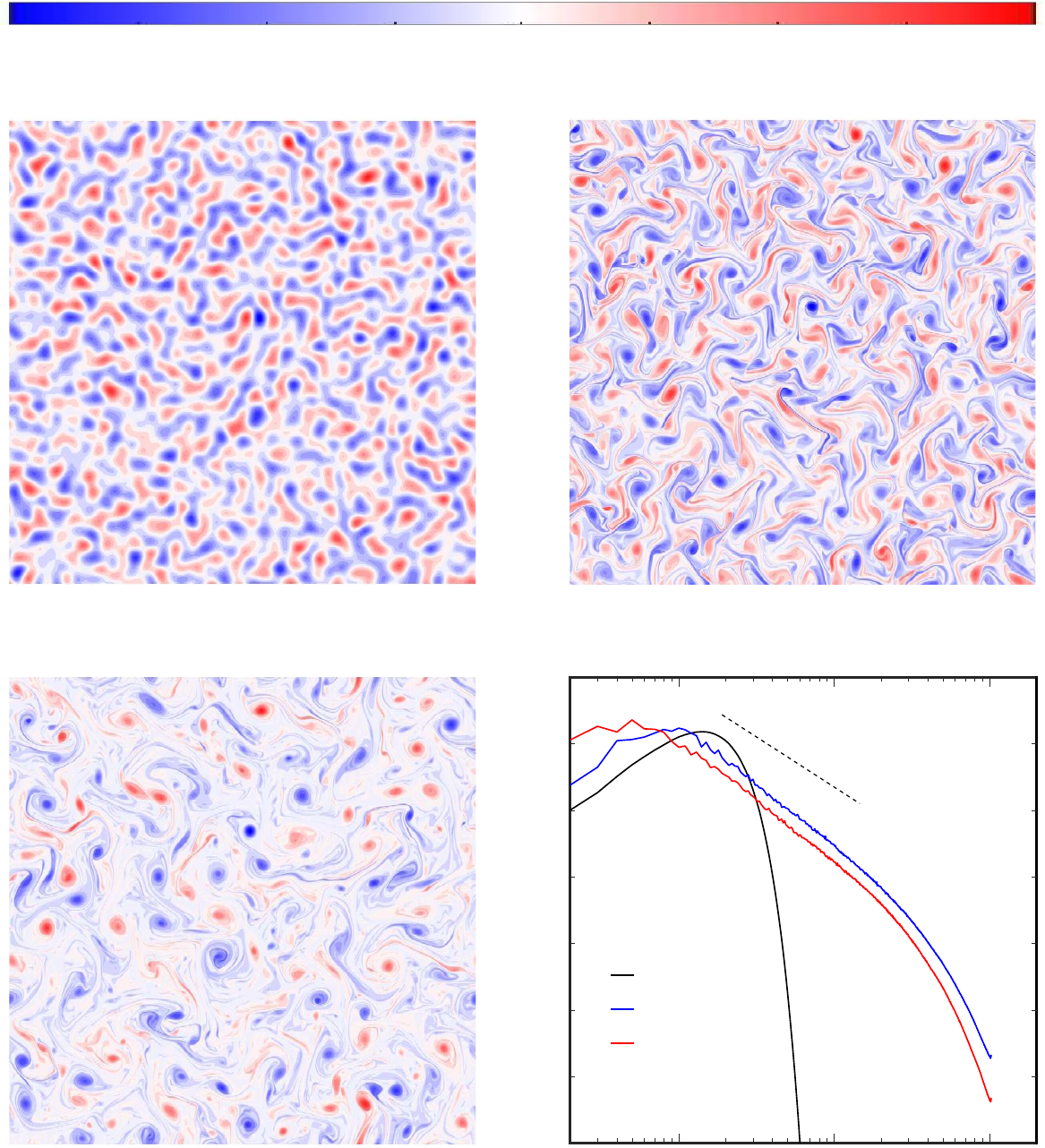}
 \put(0,95){$-50$}
 \put(23,95){$-25$}
 \put(49,95){$0$}
 \put(73,95){$25$}
 \put(97,95){$50$}

 \put(20,91){(a) t = 0}
 \put(74,91){(b) t = 50$\tau$}
 \put(19,42){(c) t = 200$\tau$}
 \put(70,42){(d) $\hat{E}(k)$ spectrum}

 \put(62,14.5){t=0}
 \put(62,11.4){t=50$\tau$}
 \put(62,8.5){t=200$\tau$}

 \put(80,33){$k^{-3}$}

 \put(50,40){\tiny $10^0$}
 \put(50,34){\tiny $10^{-2}$}
 \put(50,28){\tiny $10^{-4}$}
 \put(50,22.1){\tiny $10^{-6}$}
 \put(50,16.2){\tiny $10^{-8}$}
 \put(50,10.5){\tiny $10^{-10}$}
 \put(50,5){\tiny $10^{-12}$}
 \put(50,-0.5){\tiny $10^{-14}$}

 \put(54,-2){\tiny $10^0$}
 \put(64,-2){\tiny $10^1$}
 \put(79,-2){\tiny $10^2$}
 \put(94,-2){\tiny $10^3$}

 \put(67,-5){Wavenumber, $k$}

 \end{overpic}
  \vspace*{5mm}

 \caption{\small An example of the vorticity field of DNS for $Re = 32000$ at a) $t=0$, b) $t=50\tau$, and c) $t=200\tau$. The initial turbulent kinetic energy (TKE) spectrum is prescribed while the vorticity field has random phase (see Appendix~\ref{Initial conditions of DNS}). Data collection for training of data-driven SGS models (using CNN or ANN) starts from $t=50\tau$ and ends at $t=200\tau$. As shown in (d), in this period, the TKE spectra exhibit self-similarity following the $k^{-3}$ scaling of the Kraichnan-Batchelor-Leith (KBL) theory~\cite{kraichnan1967inertial,batchelor1969computation,leith1971atmospheric}.}
 \label{fig:1}
\end{figure}
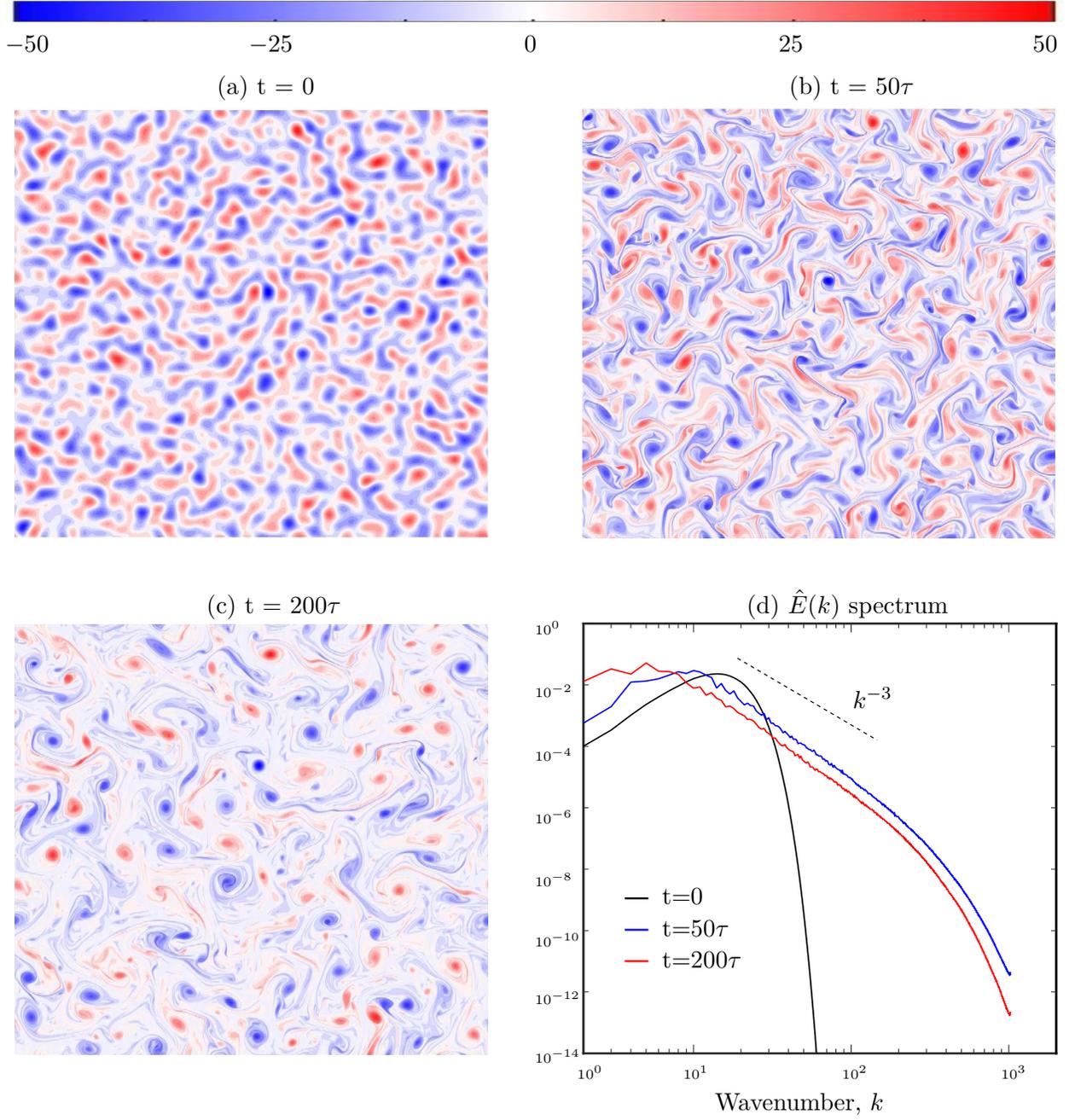

Figure~\ref{fig:1} shows an example of the vorticity field for $Re = 32000$ at the initial condition ($t=0$), and at $t = 50\tau$ and $t = 200\tau$, where $\tau=1/|\omega|_{\text{max}} = 0.02=200\Delta t_{\text{DNS}}$ ($|\omega|_{\text{max}}$ is computed at $t=0$). After around $50\tau$, the turbulent kinetic energy (TKE) spectrum ($\hat{E}(k)$) exhibits self-similarity. Note that the TKE spectrum is calculated using an angle average and therefore $k=\sqrt{k_x^2 + k_y^2}$.\\

For LES, we solve Eqs.~\eqref{FNS1}-\eqref{FNS2} using the same numerical solver used for DNS, except that the spatial resolution is lower by a factor of $8$ in each direction (i.e., $N_\text{LES}=N_\text{DNS}/8$ and $\Delta_\text{LES}=8\Delta_\text{DNS}$) and the time-stepping size is 10 times larger, $\Delta t_{\text{LES}}=10\Delta t_\mathrm{DNS}$. As a result, the LES solver requires 640 times fewer degrees of freedom, which substantially reduces the computational cost. However, the LES solver needs a SGS model for $\Pi$. Here, we use two data-driven models that employ CNN and ANN as well as two common physics-based models (SMAG and DSMAG). In the next section, we first describe the filtered DNS (FDNS) data, which  are used for training the data-driven SGS models, and then describe the CNN, ANN, SMAG, and DSMAG models.

\section{Data-driven and physics-based SGS models for LES}\label{LES methods}
\subsection{Filtered DNS (FDNS) data} \label{sec:filtering}
To compute the filtered DNS variables on the LES grid, which as mentioned above is $8\times$ coarser than the DNS grid in each direction, we i) apply the Gaussian filter transfer function to the DNS data, and ii) coarse-grain the filtered results to the LES grid \cite{pope2001turbulent,zanna2020data}. Below, the subscript ``DNS" denotes the high-resolution DNS grid and ``LES" denotes the coarse-resolution LES grid. \\

Using vorticity as an example, we first transform the DNS vorticity field $\omega(\mathbf{r}_{\text{DNS}})$ into the spectral space $\hat{\omega}(\mathbf{k_{\text{DNS}}})$, where $\mathbf{r}=(x,y)$ and $\mathbf{k}=(k_x,k_y)$. Then, we apply the Gaussian filter in the spectral space
\begin{eqnarray}
\tilde{\hat{\omega}}(\mathbf{k_{\text{DNS}}})=G(\mathbf{k_{\text{DNS}}})\odot\hat{\omega}(\mathbf{k_{\text{DNS}}}), \label{Vor_Mag1}
\end{eqnarray}
where the operator $\odot$ means element-wise multiplication of matrices and $\tilde{(\cdot)}$ denotes the filtered variable at the DNS resolution. The transfer function of the Gaussian filter is~\cite{pope2001turbulent}:
\begin{eqnarray}
G(\mathbf{k_{\text{DNS}}}) = e^{-|\mathbf{k_{\text{DNS}}}|^2\Delta_F^2/24}, \label{Vor_Mag2}
\end{eqnarray}
where $\Delta_F$ is the filter size, which is taken to be $\Delta_F = 2\Delta_{\text{LES}}$ to yield sufficient resolution~\cite{pope2001turbulent,zhou2019subgrid}. After the filtering operation, coarse-graining is performed to transform the filtered solution from the DNS to LES grid~\cite{pope2001turbulent,zanna2020data}:
\begin{eqnarray}
\overline{\hat{\omega}}(\mathbf{k_{\text{LES}}})=\tilde{\hat{\omega}}(|k_x|<k_c,|k_y|<k_c) \label{Vor_Mag3}
\end{eqnarray}
where $k_c=\pi/\Delta_{\text{LES}}$ is the cut-off wavenumber in spectral space, and we use $\overline{(\cdot)}$ to denote the filtered and then coarse-grained variables (hereafter, we use the term ``filtered" for both ``filtered" and then ``coarse-grained" when there is no ambiguity). $\overline{\hat{\psi}}(\mathbf{k_{\text{LES}}})$ and $\hat{\Pi}(\mathbf{k_{\text{LES}}})$ are similarly computed following Eqs.~(\ref{Vor_Mag1})-(\ref{Vor_Mag3}).\\

Note that in addition to the Gaussian filter, box and sharp Fourier filters are also commonly used for LES. However, the Gaussian filter is compact in both physical and spectral spaces~\cite{pope2001turbulent}. Because our numerical solver is in the Fourier spectral space and our CNN and ANN operate in the physical space, we focus on the Gaussian filter for LES. Furthermore, Zhou \textit{et al.}~\cite{zhou2019subgrid} found that the Gaussian filter outperforms the other two filters in terms of correlation coefficients of $\Pi$ in their work on data-driven SGS modeling of 3D turbulence.\\

\begin{figure}[t]
\vspace{.1in}
 \centering
 \begin{overpic}[width=1\linewidth,height=0.5\linewidth]{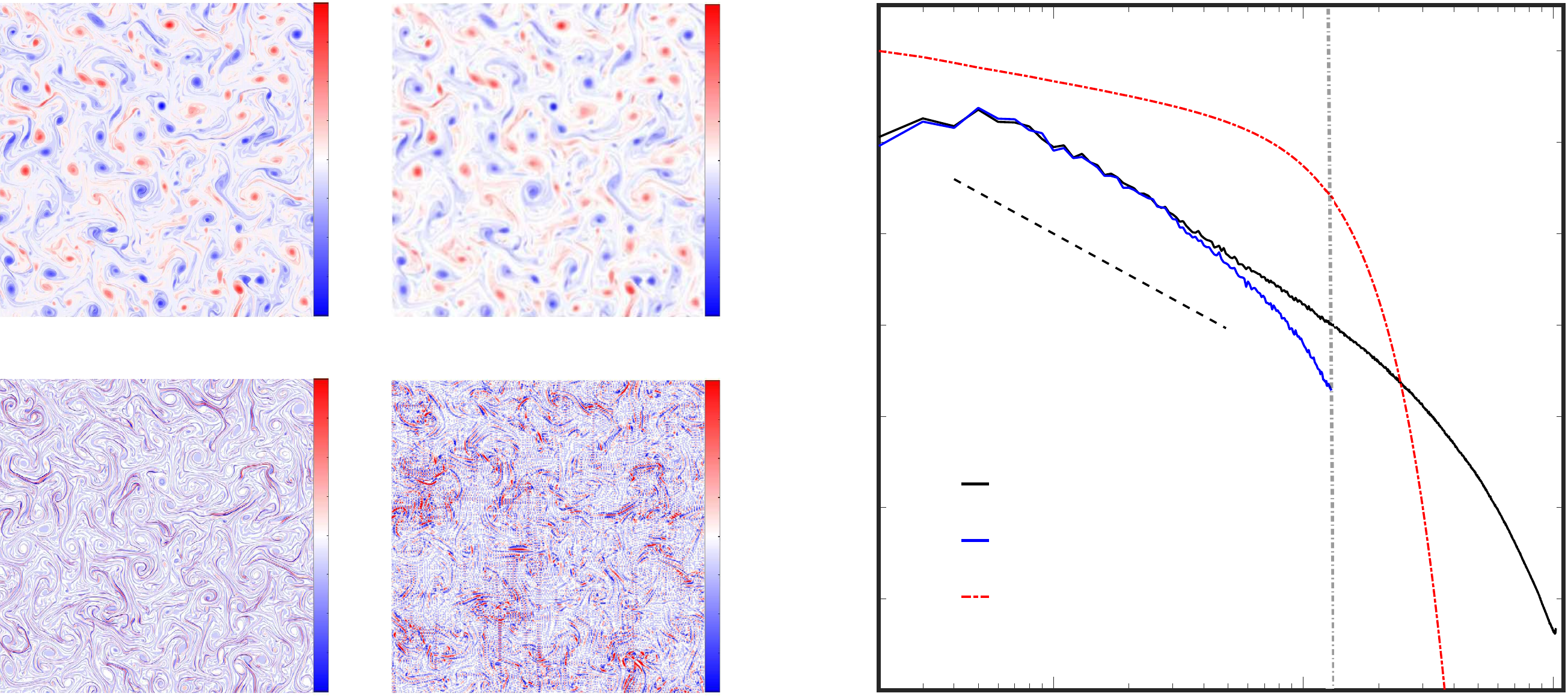}
 \put(7,51){(a) $\omega$}
 \put(32,51){(b) $\tilde{\omega}$}
 \put(58,51){(e) $\hat{E}(k)$ spectrum and Gaussian filter}
 \put(2,24){(c) $\omega' = \omega - \tilde{\omega}$}
 \put(32,24){(d) $\Pi$}

 \put(21.5,49){\tiny$50$}
 \put(21.5,38){\tiny$0$}
 \put(21,27){\tiny$-50$}

 \put(46.5,49){\tiny$50$}
 \put(46.5,38){\tiny$0$}
 \put(46,27){\tiny$-50$}

 \put(21.5,22){\tiny$10$}
 \put(21.5,11){\tiny$0$}
 \put(21,0){\tiny$-10$}

 \put(46.5,22){\tiny$10$}
 \put(46.5,11){\tiny$0$}
 \put(46,0){\tiny$-10$}

 \put(52,45.5){\tiny$10^0$}
 \put(52,38.75){\tiny$10^{-2}$}
 \put(52,32.5){\tiny$10^{-4}$}
 \put(52,26){\tiny$10^{-6}$}
 \put(52,19.5){\tiny$10^{-8}$}
 \put(51.5,13){\tiny$10^{-10}$}
 \put(51.5,6){\tiny$10^{-12}$}
 \put(51.5,0){\tiny$10^{-14}$}

 \put(55,-2){\tiny$10^0$}
 \put(66,-2){\tiny$10^1$}
 \put(82,-2){\tiny$10^2$}
 \put(98,-2){\tiny$10^3$}

 \put(65,27){$k^{-3}$}
 \put(86,2){$k_c$}

 \put(64,14.5){DNS}
 \put(64,10.5){FDNS}
 \put(64,6.0){Gaussian filter}
\put(72,-5){Wavenumber, $k$}
 \end{overpic}
 \vspace*{5mm}
 \caption{\small Examples showing the effects of filtering. a) DNS vorticity $\omega$, (b) filtered vorticity $\tilde{\omega}$, (c) SGS vorticity $\omega'=\omega-\tilde{\omega}$, (d) SGS forcing term $\Pi$, and (e) TKE spectrum for $Re = 32000$ at the end of one of the DNS runs ($t=200\tau$). Panel (e) also shows the transfer function of the Gaussian filter and the cutoff wavenumber, $k_c$. The FDNS spectrum deviates from the DNS spectrum near $k_c$ because of the filtering.}
 \label{fig:2}
\end{figure}
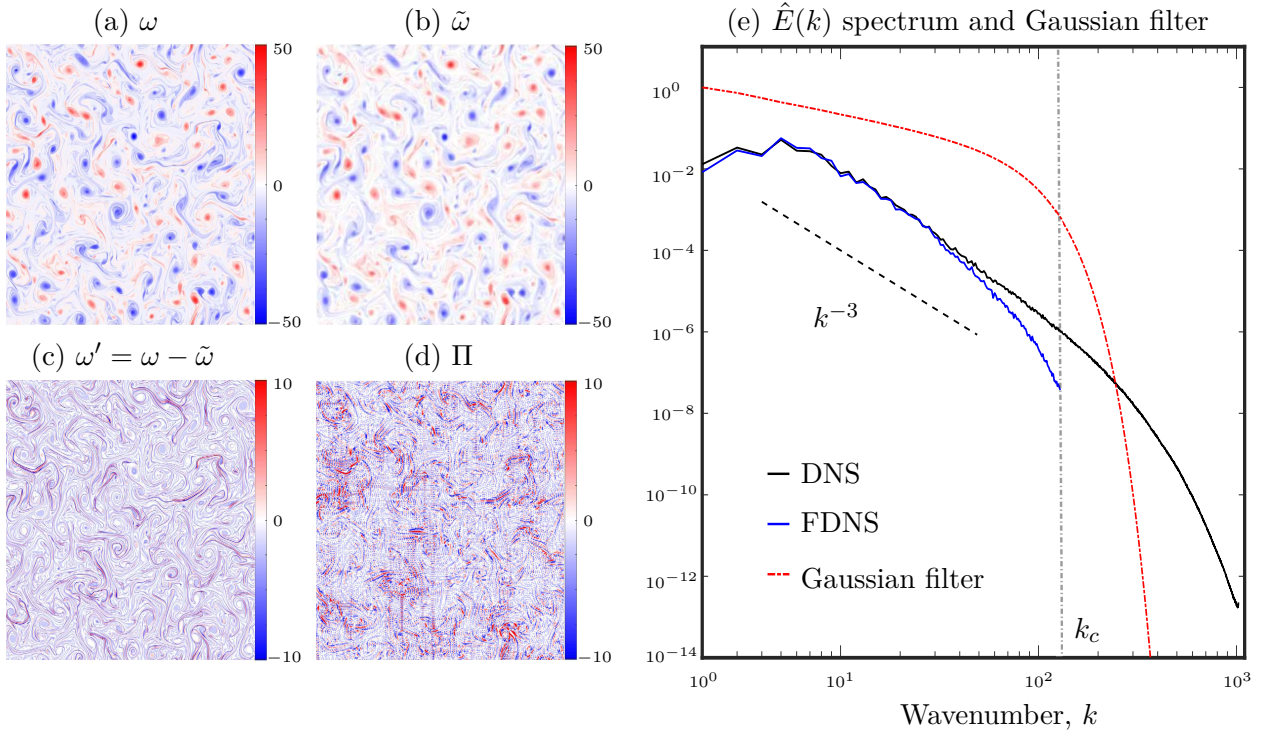

Figure~\ref{fig:2} shows examples of the $\Pi$ term and effects of filtering on the vorticity field in physical space and on the TKE spectrum ($\hat{E}(k)$). The fine structures in DNS vorticity $\omega$ are lost in filtered vorticity $\tilde{\omega}$ and manifest themselves in SGS vorticity $\omega'=\omega-\tilde{\omega}$ and the SGS forcing term $\Pi$ (panels (c)-(d)). The $\hat{E}(k)$ spectrum further shows the effects of the Gaussian filter on the energy at smaller scales (panel~(e)). The Gaussian filter leads to the deviation of the FDNS spectrum from the DNS spectrum, especially at the scales near $k_c$.\\

Our goal is to non-parameterically learn $\Pi$ as function of the FDNS variables $\overline{\omega}$ and $\overline{\psi}$ using a deep fully CNN as well as an ANN used in a previous study~\cite{maulik2019subgrid}.

\subsection{Fully convolutional neural network (CNN)} \label{sec:cnn}
For non-local data-driven SGS modeling, we propose to use a deep fully CNN. The CNN architecture was originally developed for computer vision and image processing and its key feature is that rather than having pre-defined filters, CNNs learn the filters used for pattern recognition for a given data set~\cite{lecun1990handwritten,krizhevsky2012imagenet,goodfellow2016deep}. CNNs have often been found superior to ANNs when the data contains spatial patterns and structures significant to the functional relationship to be learned~\cite{peyrard2015comparison,driss2017comparison}. Therefore, it is not surprising that CNNs have been found to perform well, usually superior to non-convolutional ML methods, in applications involving turbulent flows, given the abundance of coherent structures and spatial correlations in turbulence~\cite[e.g.,][]{bolton2019applications,mohan2019compressed,beck2019deep,chattopadhyay2020predicting,chattopadhyay2020analog,pawar2020priori}. Specifically for SGS modeling, a recent \textit{a priori} analysis has shown that CNN outperforms local ANN in terms of prediction accuracy of the SGS stress term in the same 2D-DHIT system studied here~\cite{pawar2020priori}.


Building on previous work and to account for non-local effects (e.g., coherent structures and spatial correlations), we use a CNN with inputs/outputs that are global (i.e., from the entire domain). Thus, the input features are
\begin{eqnarray}\label{CNN-input}
\Bigg\{\frac{\overline{\psi}}{\sigma_{ \overline{\psi}}},\frac{\overline{\omega}}{\sigma_{ \overline{\omega}}}\Bigg\}\in \mathbb{R}^{2\times N_{\text{LES}}\times N_{\text{LES}}},
\end{eqnarray}
and the output targets are
\begin{eqnarray}\label{CNN-output}
\Bigg\{\frac{\Pi}{\sigma_{ \Pi}}\Bigg\}\in \mathbb{R}^{N_{\text{LES}}\times N_{\text{LES}}},
\end{eqnarray}
where $\sigma$ is the standard deviation of the corresponding variables calculated over all training samples. We aim to use a CNN to learn $\mathbb{M}$, an optimal map between the inputs and outputs
\begin{eqnarray}\label{CNN-input}
\mathbb{M}:\Big\{\overline{\psi}/\sigma_{\overline{\psi}},\overline{\omega}/\sigma_{\overline{\omega}}\Big\} \in \mathbb{R}^{2\times N_{\text{LES}}\times N_{\text{LES}}} \rightarrow \Big\{\Pi/\sigma_{\Pi}\Big\} \in \mathbb{R}^{\times N_{\text{LES}}\times N_{\text{LES}}}
\end{eqnarray}
by minimizing the mean-squared-error ($MSE$)
\begin{eqnarray}
MSE = \frac{1}{n_{tr}}\sum_{i=1}^{n_{tr}}\parallel \Pi_i^{\text{CNN}}-\Pi_i^{\text{FDNS}}\parallel_2^2, \label{eq:mse}
\end{eqnarray}
where $n_{tr}$ is the number of training samples and $\parallel \cdot \parallel_2$ is the $L_2$ norm.\\

\begin{figure}[t]
\vspace{.1in}
 \centering
 \begin{overpic}[width=0.95\linewidth,height=0.12\linewidth]{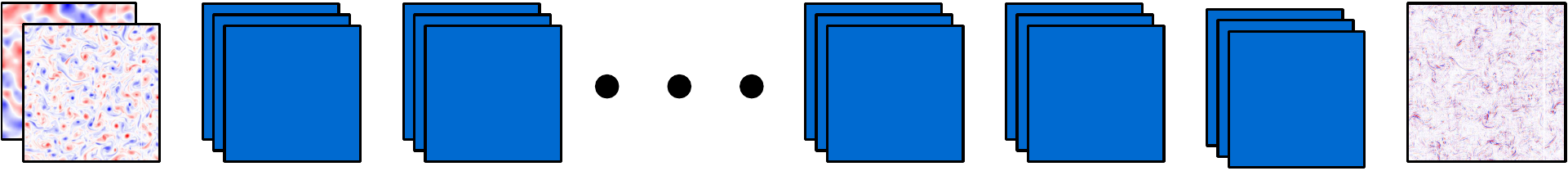}
 \put(-2,0){$\bar{\psi}$}
 \put(1,-2){$\bar{\omega}$}
 \put(94,-2){$\Pi$}

 \put(16.0,6){\textcolor{white}{\small Conv}}
 \put(16.0,3){\textcolor{white}{\small layer}}
  \put(28.7,6){\textcolor{white}{\small Conv}}
 \put(28.7,3){\textcolor{white}{\small layer}}
  \put(54.3,6){\textcolor{white}{\small Conv}}
 \put(54.3,3){\textcolor{white}{\small layer}}
  \put(67.2,6){\textcolor{white}{\small Conv}}
 \put(67.2,3){\textcolor{white}{\small layer}}
 \put(80,6){\textcolor{white}{\small Conv}}
 \put(80,3){\textcolor{white}{\small layer}}

 \end{overpic}
 \vspace*{5mm}
 \caption{\small Schematic of the CNN. Inputs and outputs are samples of normalized $(\bar{\psi},\bar{\omega})$ and $\Pi$, respectively. The convolutional layers (Conv layers) have the same dimension ($256 \times 256$) as that of the the input and output layers. All Conv layers are initialized randomly and are trainable. The convolutional depth is set to 64, and the convolutional filter size is $5\times 5$. The activation function of each layer is ReLu (rectified linear unit) except for the last one, which is a linear map.}
 \label{fig:3}
\end{figure}
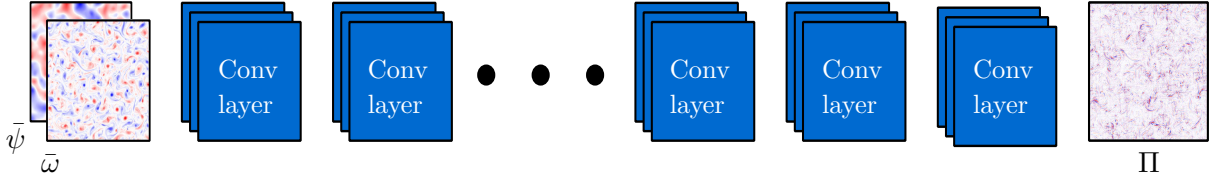

Figure~\ref{fig:3} schematically shows the CNN architecture that is used here. We use the mini-batch stochastic gradient descent method with the Adam optimizer to minimize the loss function, Eq.~\eqref{eq:mse}. Note that the CNN has no pooling or upsampling layers (i.e., fully CNN), so the hidden layers have the same size as the input and output layers. We have found that using a fully CNN (i.e. without an up/down sampling) is a key to training an accurate SGS model, consistent with earlier findings that pooling layers may artificially change spatial correlations of the data~\cite{chattopadhyay2020analog}.\\

Hyper-parameters such as the number of hidden layers have been determined via an extensive search. We find that to capture the complex pattern of $\Pi$, a deep CNN with 10 hidden layers is needed. For example, the 10-layer CNN outperforms shallower 8-layer and 5-layer CNNs in terms of training loss for the same $n_{tr}$. Overall, the CNN with 10 layers has $927041$ trainable parameters.\\

The training, validation, and testing sets are generated using 2D snapshots of filtered data collected from $15$ independent DNS runs with random initial conditions, sampled every $10\Delta_\text{DNS}$, in the time interval $[50\tau,200\tau]$. We use the data from $8$ runs for the training set, $2$ runs for the validation set, and $5$ runs for the testing set. The effects of the size of the training dataset on the accuracy of the SGS model is further discussed in Section~\ref{a priori analysis}.\\

As is the common practice in ML applications, we run the CNN (and the ANN) with single-precision floating-point operations during both training and testing to accelerate the process and reduce the data transfer/storage. We have also explored training/testing a CNN with double-precision floating-point arithmetic, but found no distinguishable enhancement in the \textit{a posteriori} tests.\\

Finally, we point out that the codes for CNN and CNN with transfer learning (discussed later) are made publicly available on GitHub (see the Acknowledgement for details).

\subsection{Multilayer perceptron artificial neural network (ANN)} \label{sec:ann}
A few recent studies have proposed building local data-driven SGS models using ANNs trained to learn the mapping between a local stencil of input variables to the local SGS term $\Pi$~\cite{maulik2019subgrid,xie2019artificial,zhou2019subgrid,xie2021artificial}. For example, Maulik~\textit{et al.}~\cite{maulik2019subgrid} employed such an approach for the same 2D-DHIT system and proposed to train an ANN with inputs consisting of 9 grid stencil values of $\overline{\omega}$ and $\overline{\psi}$ plus the local values of $|\overline{S}|$ and $|\nabla\overline{\omega}|$ and the output consisting of the local SGS term $\Pi$ value:
\begin{eqnarray}\label{ANN-map}
\mathbb{M}:\Big\{\overline{\omega}_{i,j},\overline{\omega}_{i,j+1},...,\overline{\omega}_{i-1,j-1},\overline{\psi}_{i,j},\overline{\psi}_{i,j+1},
...,\overline{\psi}_{i-1,j-1},|\overline{S}|_{i,j},|\nabla\overline{\omega}|_{i,j}\Big\}\in \mathbb{R}^{20} \rightarrow \Big\{\Pi_{i,j}\Big\}\in \mathbb{R}^{1},
\end{eqnarray}
where $(i,j)$ here denotes a local grid point. $|\overline{S}|$ is the characteristic filtered rate of strain \citep{pope2001turbulent} and $|\nabla\overline{\omega}| = \sqrt{\left(\frac{\partial \overline{\omega}}{\partial x}\right)^2+\left(\frac{\partial \overline{\omega}}{\partial y}\right)^2}$.\\

We have closely followed Ref.~\cite{maulik2019subgrid} in building a local data-driven SGS model. For the ANN, we use their publicly available code. The ANN is fully connected with 2 hidden layers, each containing 50 neurons. The network has 3651 trainable parameters. We explore architectures with more layers and neurons per layer, but find no improvement in the accuracy. Due to the use of local inputs, in this approach the number of training samples is equal to the number of snapshots multiplied by $N^2_\text{LES}$. In common practice, only a few (less than 10) snapshots of data is used as the training data set~\cite{maulik2019subgrid,zhou2019subgrid,xie2019artificial,xie2020modeling}. Here, following Ref.~\cite{maulik2019subgrid}, we use 8 randomly selected snapshots (from the training set mentioned in Section~\ref{sec:cnn}) resulting in $524288$ samples in the training sets. We have also investigated the effects of increasing the number of samples to 20 snapshots, but again, no substantial improvement in training loss is found. Note that following Ref.~\cite{maulik2019subgrid}, no pre-processing, e.g., normalization, is performed on the input or output data (we find normalizing the input/output samples to have no effect on the performance of the ANN).\\

Note that it is not the purpose of this paper to compare the ANN- and CNN-based approaches side by side (even if such comparison is possible given the differences in architecture, network size, input/output, and size of the training set). Therefore, beyond the explorations mentioned above, we have not performed an exhaustive search on the ANN and local SGS modeling approach. Our explorations all suggest that the comprehensively investigated network/approach presented in Ref.~\cite{maulik2019subgrid} is already optimal.

\subsection{Smagorinsky (SMAG) and dynamics Smagorinsky (DSMAG) SGS models} \label{sec:smag}
In the SMAG~\cite{smagorinsky1963general} model, which is a commonly used baseline SGS model for LES, the SGS stress term in the momentum equation is modeled as~\cite{pope2001turbulent,sagaut2006large}:
\begin{eqnarray}\label{SMAG-stress}
\boldsymbol{\tau}^{\text{SMAG}}= -2(C_s\Delta)^2\langle 2\overline{S}\:\overline{S}\rangle^{1/2}\overline{S},
\end{eqnarray}
where the angle brackets $\langle\cdot\rangle$ denote domain averaging. $\overline{S}$ is the filtered rate-of-strain tensor \citep{pope2001turbulent}. The SGS term $\Pi$ in Eq.~\eqref{FNS1} is therefore:
\begin{eqnarray}\label{SMAG}
\Pi^{\text{SMAG}} = (C_s\Delta)^2\langle 2\overline{S}\:\overline{S}\rangle^{1/2}\nabla^2\overline{\omega} = \nu_e\nabla^2\overline{\omega},
\end{eqnarray}
where $C_s$ is the Smagorinsky coefficient, $\nu_e$ is the eddy viscosity, and
\begin{eqnarray}\label{strain-rate-mag}
\langle 2\overline{S}\:\overline{S}\rangle^{1/2} = \sqrt{4\left(\frac{\partial^2\overline{\psi}}{\partial x\partial y}\right)^2 +\left(\frac{\partial^2\overline{\psi}}{\partial x^2} - \frac{\partial^2\overline{\psi}}{\partial y^2}\right)^2}.
\end{eqnarray}
$C_s$ is a constant in the SMAG model. The DSMAG model \cite{germano1991dynamic} uses a dynamic procedure to estimate $\nu_e$ based on the local flow structure. This procedure can lead to $\nu_e<0$, which can result in numerical stabilities; consequently, ``positive clipping'' is often applied to enforce $\nu_e \ge 0$~\cite{zang1993dynamic}. Here, we use $C_s=1$ for SMAG following Maulik~\textit{et al.}~\cite{maulik2019subgrid} and implement DSMAG (with positive clipping) following Pawar~\textit{et al.}~\cite{pawar2020priori}, who studied the same 2D-DHIT system. Note that these SMAG and DSMAG models both have $\nu_e \ge 0$ and are therefore purely diffusive.


\section{Results}\label{sec:results}
\subsection{\textit{A priori} analysis}\label{a priori analysis}
\subsubsection{Accuracy} \label{sec:accuracy}
We first examine the accuracy of the CNN-based SGS model in predicting the $\Pi$ term and inter-scale transfers for never-seen-before samples of $(\overline{\psi},\overline{\omega}$) from the testing set. The results in Section~\ref{sec:accuracy} are reported for $n_{tr}=50000$. We use a commonly used metric, the correlation coefficient $c$ between the modeled ($\Pi^{M}$) and true ($\Pi^{\text{FDNS}}$) SGS terms defined as ~\cite{zhou1991eddy,beck2019deep,pawar2020priori}:
\begin{eqnarray}\label{cc}
c = \frac{\Big\langle\Big(\Pi^{M} - \langle\Pi^{M}\rangle\Big) \Big( \Pi^{\text{FDNS}} - \langle\Pi^{\text{FDNS}}\rangle\Big)\Big\rangle}{\sqrt{\Big\langle(\Pi^{M} - \langle\Pi^{M}\rangle)^2\Big\rangle}\sqrt{\Big\langle(\Pi^{\text{FDNS}} - \langle\Pi^{\text{FDNS}}\rangle)^2\Big\rangle}}.
\end{eqnarray}
The correlation coefficients (averaged over 100 random testing samples) for CNN as well as DSMAG and ANN are reported in Table~\ref{table1}. These \textit{a priori} tests show that the data-driven SGS models substantially outperform DSMAG, and that this CNN-based model (with $c$ above $0.9$) has statistically significantly higher accuracy than this ANN-based model. Note that similarly, previous findings based on correlation coefficients of SGS stress term found CNNs to outperform ANNs in \textit{a priori} tests~\cite{pawar2020priori}. \\

\begin{table}[tb]
\centering
\caption{Correlation coefficients $c$ (Eq.~\eqref{cc}) between the predicted and true SGS term $\Pi$ for $Re = 32000$ in \textit{a priori} tests. The subscripts indicate $c$ computed only over elements of $\Pi^{FDNS}$ and $\Pi^{M}$ corresponding to $T>0$ or $T<0$ (Eq.~\eqref{SGStransfer}). The values show the average over (the same) $100$ randomly chosen testing samples and the standard deviation.}\label{table1}
\begin{tabular}{|l|l|l|l|}
\cline{1-4}
&DSMAG&ANN&CNN\\
\cline{1-4}\hline\hline
$c$&$0.55\pm0.06$&$0.86\pm0.02$&$0.93\pm0.03$\\
\cline{1-4}
$c_{T>0}$&$0.55\pm0.06$&$0.86\pm0.02$&$0.96\pm0.03$\\
\cline{1-4}
$c_{T<0}$&$0$&$0.83\pm0.02$&$0.92\pm0.04$\\
\cline{1-4}
\end{tabular}
\label{table1}
\end{table}

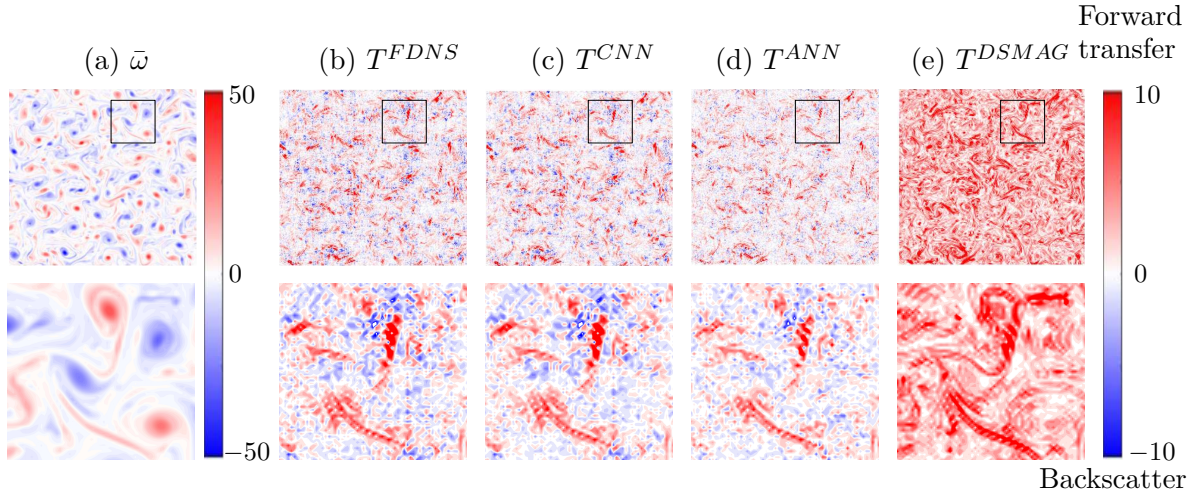
\begin{figure}[t]
\vspace{.1in}
\vspace*{15mm}
 \centering
 \begin{overpic}[width=0.9\linewidth,height=0.3\linewidth]{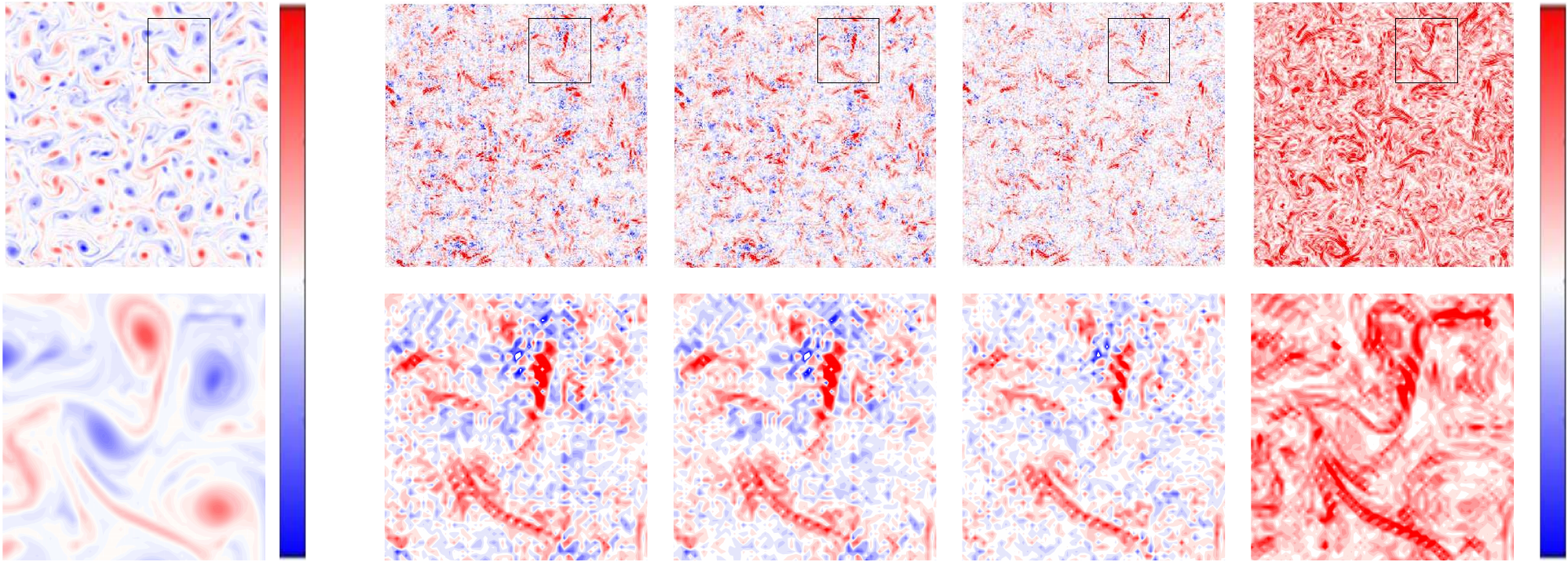}
 \put(7,35){(a) $\bar{\omega}$}
 \put(28,35){(b) $T^{FDNS}$}
 \put(47.0,35){(c) $T^{CNN}$}
 \put(63.5,35){(d) $T^{ANN}$}
 \put(81,35){(e) $T^{DSMAG}$}

 \put(92.5,-2.5){Backscatter}
 \put(96,39){Forward}
 \put(96,36){transfer}

 \put(20,32){\small $50$}
 \put(20,16){\small $0$}
 \put(19.5,0){\small $-50$}

 \put(101,32){\small $10$}
 \put(101,16){\small $0$}
 \put(100.5,0){\small $-10$}

 \end{overpic}
  \vspace*{5mm}
 \caption{\small Example of inter-scale transfer $T$, Eq.~\eqref{SGStransfer}, in \textit{a priori} analysis at $Re = 32000$. a) Filtered vorticity $\overline{\omega}$; b) true $T$ from FDNS; (c)-(e) $T$ from CNN, ANN, and DSMAG. The ANN and CNN capture both forward transfer and backscatter while DSMAG only captures the forward transfer (diffusion). The upper row shows the entire domain while the second row shows the portion in the black square.}
 \label{fig:5}
\end{figure}

Next, we examine the inter-scale transfer in \textit{a priori} tests. The transfer is often quantified using the SGS stress~\cite{pope2001turbulent}. Since here we are working with the SGS forcing term $\Pi$, which is the curl of the divergence of the SGS stress, we instead follow previous work and define SGS transfer $T$ as~\cite{kerr1996small,thuburn2014cascades,maulik2019subgrid}:
\begin{eqnarray}\label{SGStransfer}
T = sgn(\nabla^{2}\overline\omega)\odot \Pi,
\end{eqnarray}
where $sgn(\cdot)$ is the sign function. At each grid point ($i,j$), $T_{i,j}>0$ indicates forward transfer (diffusion) while $T_{i,j}<0$ indicates backscatter. Note that forward/backscatter is between the resolved and subgrid scales as separated by filtering, and should not be confused with the forward/inverse cascade, which is a physical property~\cite{pope2001turbulent,vallis2017atmospheric}. For a sample filtered vorticity $\overline{\omega}$, Fig.~\ref{fig:5} shows the true inter-scale transfer $T^\text{FDNS}$ and $T$ from CNN, ANN, and DSMAG. Because DSMAG is purely diffusive, it only captures the forward transfer. The ANN and CNN both capture the diffusion as well as backscattering. Table~\ref{table1} further shows $c$ computed separately over grid points corresponding to only $T>0$ (diffusion) or only $T<0$ (backscattering), again, demonstrating that the CNN-based SGS model captures both forward transfer of backscatter accurately, with $c>0.9$.\\

To summarize, the \textit{a priori} tests show that the CNN-based data-driven SGS model can accurately predict the out-of-sample SGS forcing terms and inter-scale transfers. However, as discussed in the Introduction, previous studies have found that accuracy in \textit{a priori} tests does not necessarily translate to accuracy/stability in \textit{a posteriori} analysis~\cite{maulik2019subgrid,beck2019deep,zhou2019subgrid,zanna2020data}. Before discussing the \textit{a posteriori} tests in Section~\ref{sec:post}, we further examine how the accuracy of the CNN depends on the size of the training set, which as it turns out, impacts the stability of LES-CNN.


\subsubsection{Scaling of the CNN's accuracy with size of the training set $n_{tr}$}
Table~\ref{table2} shows how the SGS term's correlation coefficient $c$ varies in \textit{a priori} tests as the number of samples used to train the CNN ($n_{tr}$) is increased. The value of $c$ increases with $n_{tr}$, reaching $0.90$ with $n_{tr}=10000$ and $0.93$ with $n_{tr}=50000$. While $c= 0.90$ (for $n_{tr}=10000$) might seem high enough and the CNN-based data-driven SGS model might seem accurate enough, a set of \textit{a posteriori} tests with this LES-CNN model are found to lead to noisy, unphysical flows for some initial conditions. In fact, \textit{a posteriori} tests with LES-CNN trained with lower $n_{tr}$ (500 or 1000) lead to numerically unstable simulations that blow-up. Only simulations with $n_{tr} \ge 30000$ are found to lead to stable and accurate a posteriori LES-CNN for any initial condition.  \\

The above analysis suggests that instabilities in \textit{a posteriori} tests might be due to inaccurate out-of-sample predictions as a result of insufficient training data. These findings are consistent with our recent work on data-driven SGS modeling of forced 1D Burgers' turbulence with a non-local ANN~\cite{subel2020data}, where we found unstable \textit{a posteriori} LES-ANN, which was traced to inaccurate $\Pi$ terms around some of the shockwaves. In that study, we showed that artificially enriching the training dataset using a data augmentation strategy \cite{xie2018tempogan,pan2018long,formentin2019nonlinear} led to a stable and accurate LES-ANN.\\

Table~\ref{table2} further reports $c_{T>0}$ and $c_{T<0}$ as a function of $n_{tr}$. This analysis shows that consistently, $c_{T<0}$ is lower than $c_{T>0}$, especially at small $n_{tr}$, but the difference declines from $0.15$ to $0.04$ with increasing $n_{tr}$. The implication of these results is that the SGS model with a CNN trained using a small $n_{tr}$ is less capable of accurately predicting backscattering than forward transfer, which, based on previous findings, could lead to instabilities. As discussed in the Introduction, capturing backscattering is highly desired; however, it is known from physics-based SGS modeling efforts that it can lead to instabilities if handled incorrectly~\cite{lilly1992proposed,meneveau1997dynamic}. Moreover, in recent data-driven SGS modeling efforts, as discussed later, removing backscattering has been used as a way of stabilizing \textit{a posteriori} LES~\cite{maulik2019subgrid,zhou2019subgrid}. Table~\ref{table2} shows that at least for our CNN, the backscattering can be accurately captured and the \textit{a posteriori} LES can be stable without any further post-processing if the training set is large enough.    \\

In short, these results suggest that neural networks that may ``seem'' well-trained and accurate in \textit{a priori} (offline) tests, may not be sufficient for stable/accurate LES in \textit{a posteriori} (online) tests. We say ``seem'' because there is no established \textit{a priori} metric and threshold to know if a data-driven SGS model is well-trained and accurate enough to lead to stable and accurate \textit{a posteriori} LES. In this study, the threshold is empirically between $c=0.90$ and $c=0.92$, or if $c_{T<0}$ is a better metric, between $0.89$ and $0.91$. To be clear, these are just empirical thresholds in this testcase, and such thresholds might be case-dependent. Whether a general connection between a data-driven SGS models' accuracy in \textit{a priori} tests and the \textit{a posteriori} LES stability could be established or not should be thoroughly investigated in future work. Furthermore, we emphasize that we do not claim that all instabilities in other \textit{a posteriori} LES runs using data-driven SGS models (reported in other studies) are due to similar inaccuracies that could be reduced by enriching the training set.

\begin{table}[bt]
\centering
\caption{Correlation coefficients $c$ (Eq.~\eqref{cc}) between the CNN-predicted and true SGS term $\Pi$ for $Re = 32000$ in \textit{a priori} tests as a function of the number of training samples $n_{tr}$. The values show the average over (the same) $100$ randomly chosen testing samples and the standard deviation. The last row indicates the fate of \textit{a posteriori} LES-CNN integrations from 5 random initial conditions: unstable refers to numerical blow-up, unphysical refers to simulations leading to noisy/unrealistic flows, and stable refers to numerically stable and accurate simulations.}\label{table2}
\begin{tabular}{|l|l|l|l|l|l|}
\cline{1-6}
$n_{tr}$&500&1000&10000&30000&50000\\
\cline{1-6}\hline\hline
$c$&$0.78\pm 0.05$&$0.83\pm 0.04 $&$0.90\pm 0.04 $&$0.92\pm 0.04 $&$0.93\pm 0.03 $\\
\cline{1-6}
$c_{T>0}$&$0.78\pm 0.05$&$0.86\pm 0.03 $&$0.93\pm 0.04 $&$0.95\pm 0.04 $&$0.96\pm 0.03 $\\
\cline{1-6}
$c_{T<0}$&$0.63\pm 0.04$&$0.76\pm 0.03 $&$0.89\pm 0.04 $&$0.91\pm 0.04 $&$0.92\pm 0.04 $\\
\cline{1-6}\hline\hline
     &unstable&unstable&unphysical&stable&stable\\
\cline{1-6}

\end{tabular}
\label{table2}
\end{table}

\subsection{\textit{A posteriori} analysis} \label{sec:post}
In the \textit{a posteriori} (online) tests, the CNN-based data-driven SGS model and the LES numerical solver of Eqs.~\eqref{FNS1}-\eqref{FNS2} are coupled (LES-CNN): at a given time step, the resolved variables $(\overline{\psi},\overline{\omega})$ from the numerical solver are normalized (dividing by their $\sigma$) and fed into the already trained CNN, which predicts $\Pi^\text{CNN}$. This $\Pi^\text{CNN}$ is then de-normalized (multiplying by $\sigma_\Pi$) and fed back into the numerical solver to compute the resolved flow in the next time step, and the cycle continues. The CNN used for the \textit{a posteriori} tests is trained with $n_{tr}=50000$ and leads to stable LES-CNN in all tests conducted here. Similarly, we use the ANN-based data-driven SGS model and the physics-based SGS model SMAG and DSMAG to conduct LES-ANN, LES-SMAG, and LES-DSMAG integrations. \\

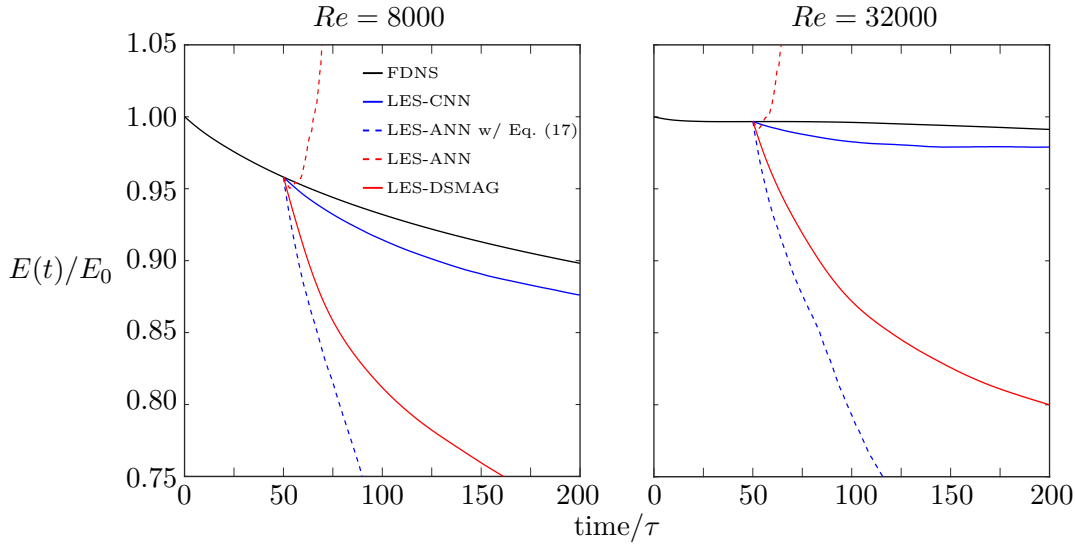
\begin{figure}[t]
\vspace{.1in}
\vspace*{15mm}
 \centering
 \begin{overpic}[width=0.7\linewidth,height=0.35\linewidth]{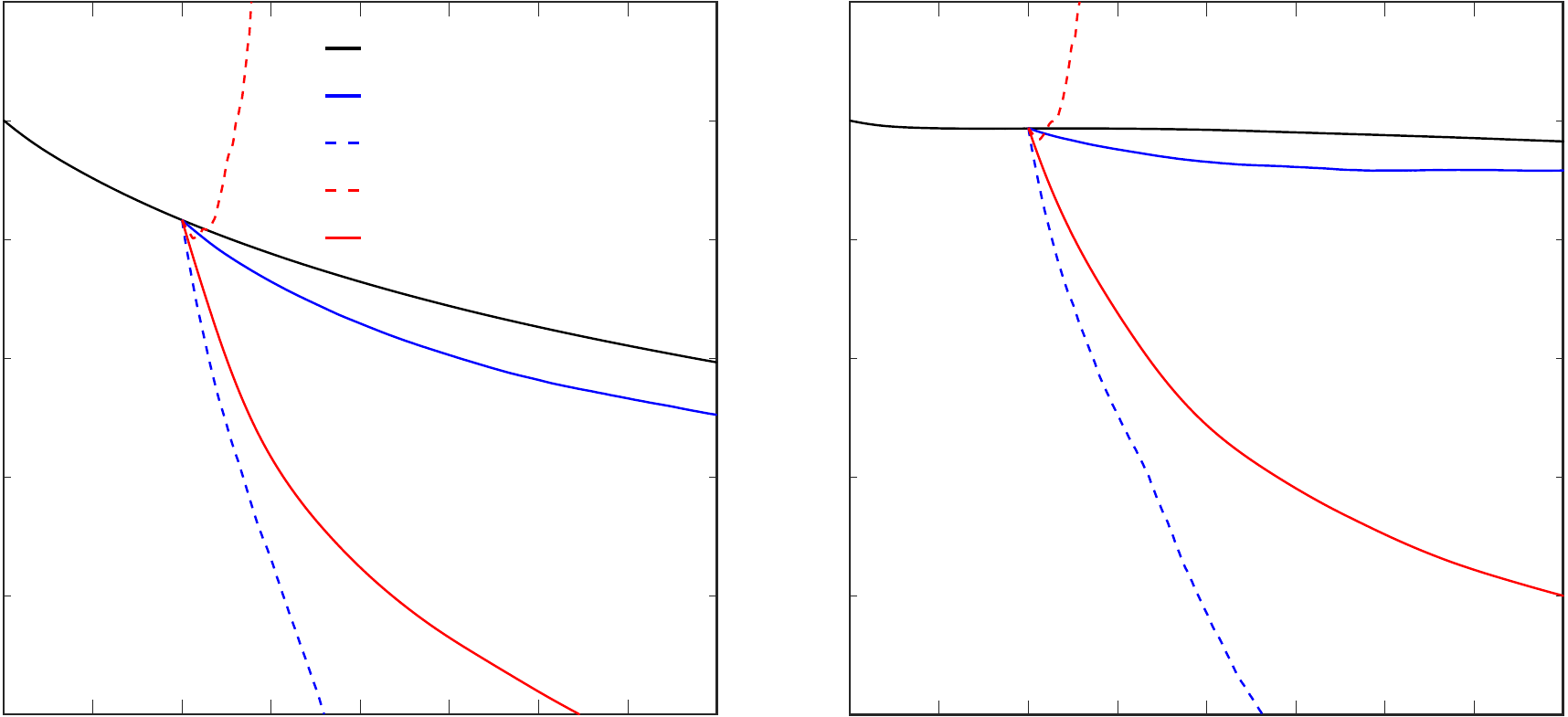}

 \put(15,52){$Re = 8000$}
 \put(70,52){$Re = 32000$}

 \put(23.5,46){\tiny FDNS}
 \put(23.5,42.8){\tiny LES-CNN}
 \put(23.5,39.5){\tiny LES-ANN w/ Eq.~\eqref{clip}}
 \put(23.5,36.2){\tiny LES-ANN}
 \put(23.5,32.8){\tiny LES-DSMAG}

 \put(-0.6,-3){$0$}
 \put(10,-3){$50$}
 \put(20.5,-3){$100$}
 \put(32,-3){$150$}
 \put(43,-3){$200$}

 \put(53.5,-3){$0$}
 \put(64,-3){$50$}
 \put(75.0,-3){$100$}
 \put(86,-3){$150$}
 \put(97.5,-3){$200$}

 \put(45,-6.5){time/$\tau$}

 \put(-6.5,-1){0.75}
 \put(-6.5,8){0.80}
 \put(-6.5,16){0.85}
 \put(-6.5,24.2){0.90}
 \put(-6.5,32){0.95}
 \put(-6.5,40.5){1.00}
 \put(-6.5,49){1.05}
 \put(-20,23){$E(t)/E_0$}

 \end{overpic}
 \vspace*{5mm}
 \caption{\small Evolution of kinetic energy $E(t)$ normalized by $E_0=E(0)$ in \textit{a posteriori} tests from 5 random initial conditions at $Re=8000$ and $Re=32000$. Note that for each $Re$, the ANN- and CNN-based data-driven SGS models have been trained on data from that $Re$. Curves show the mean from the $5$ integrations. The LES integrations start at $t=50\tau$. All stable LES models overpredict the decay rate but LES-CNN is closest to the FDNS while LES-DSMAG, and even more so the post-processed LES-ANN with backscattering removed, are too dissipative. LES-ANN without post-processing is unstable and blows up.}
 \label{fig:9}
\end{figure}

Figure~\ref{fig:9} shows examples of the evolution of the kinetic energy $E(t)=-\langle \overline{\psi} \overline{\omega} \rangle/2$ of the 2D-DHIT flow from FDNS and from the different LES models for $Re=32000$ as well as for $Re=8000$. While the LES-CNN and LES-DSMAG are stable, LES-ANN is unstable, leading to rapid increases in $E$ and blow up. In their pioneering work, Maulik~\textit{et al.}~\cite{maulik2019subgrid} also found this LES-ANN unstable and proposed a post-processing step:
\begin{eqnarray}\label{clip}
\Pi^\text{ANN}_{i, j} = 0, \text{\:\;}\forall \text{\:\;} T_{i, j}<0,
\end{eqnarray}
which effectively, like the positive clipping used for DSMAG, eliminates backscattering based on $T$ from Eq.~\eqref{SGStransfer}. A similar procedure was used by Zhou~\textit{et al.}~\cite{zhou2019subgrid} to stabilize their LES-ANN for 3D-DHIT. While this post-processed LES-ANN is stable, it is excessively dissipative (even more than DSMAG) and substantially overpredicts the energy decay rate. LES-CNN, which is stable without any post-processing and accounts for both diffusion and backscattering, has the closest agreement with FDNS in terms of the decay rate. It should be pointed out that it is possible that increasing the number of training samples for the ANN also leads to a more accurate and perhaps a stable LES-ANN; however, as mentioned before, the focus of this work is on LES-CNN and a comprehensive investigation of LES-ANN is beyond the scope of this paper. We present the results with LES-DSMAG as a baseline and present the results with the recently published LES-ANN to give the readers a better view of the state-of-the-art in this field. \\

\begin{figure}[t]
\vspace{.1in}
 \centering
 \begin{overpic}[width=0.7\linewidth,height=0.35\linewidth]{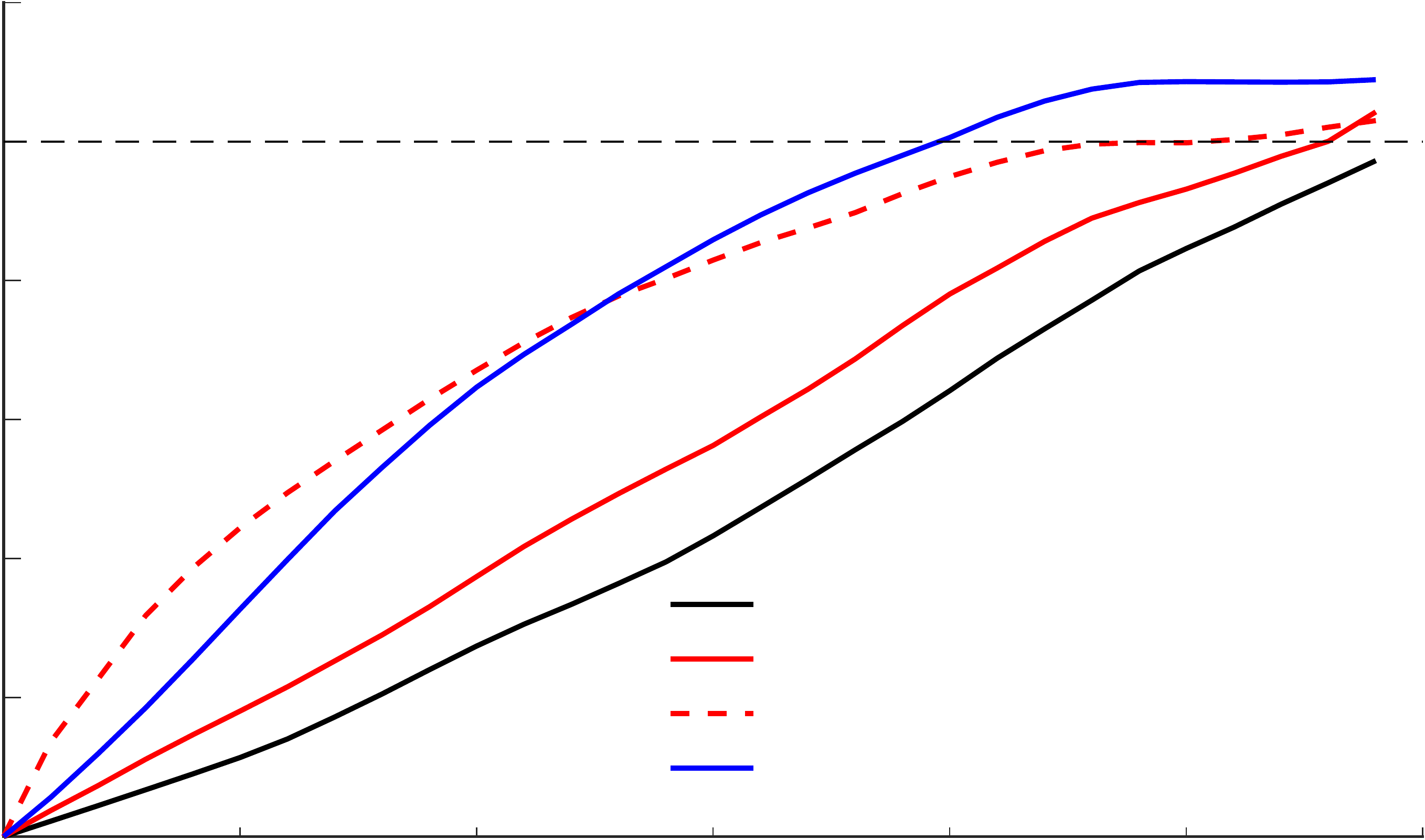}
 \put(54,13.5){\footnotesize LES-CNN}
 \put(54,10){\footnotesize LES-DSMAG}
 \put(54,6.5){\footnotesize LES-SMAG}
 \put(54,3){\footnotesize LES-ANN w/ Eq.~\eqref{clip}}

 \put(-16,50){\small Relative}
 \put(-16,46){\small  error:}
 \put(-16,42){\small $error_{L_2}$}
 \put(44,-6){time/$\tau$}

 \put(-3.5,-1){0}
 \put(-4.5,7.5){0.2}
 \put(-4.5,16){0.4}
 \put(-4.5,24.1){0.6}
 \put(-4.5,32){0.8}
 \put(-4.5,41.0){1.0}
 \put(-1,-3){50}
 \put(31,-3){100}
 \put(64,-3){150}
 \put(97,-3){200}


 \end{overpic}
 \vspace*{10mm}
 \caption{\small Short-term prediction accuracy of LES models in \textit{a posteriori} tests at $Re = 32000$. Predictions start at $t=50\tau$ in the 5 testing sets. For each model, curves show the evolution of the relative $L_2$-norm error, $error_{L_2}(t)= {\parallel \bar{\omega}^{\text{LES}}-\bar{\omega}^{\text{FDNS}}\parallel_2}/{\parallel \bar{\omega}^{\text{FDNS}}\parallel_2}$, averaged over the 5 integrations.  The LES-CNN has the highest accuracy and outperforms LES-DSMAG. The large error in the post-processed LES-ANN and in LES-SMAG is due to excessive dissipation (see Fig.~\ref{fig:6}).}
 \label{fig:7}
\end{figure}

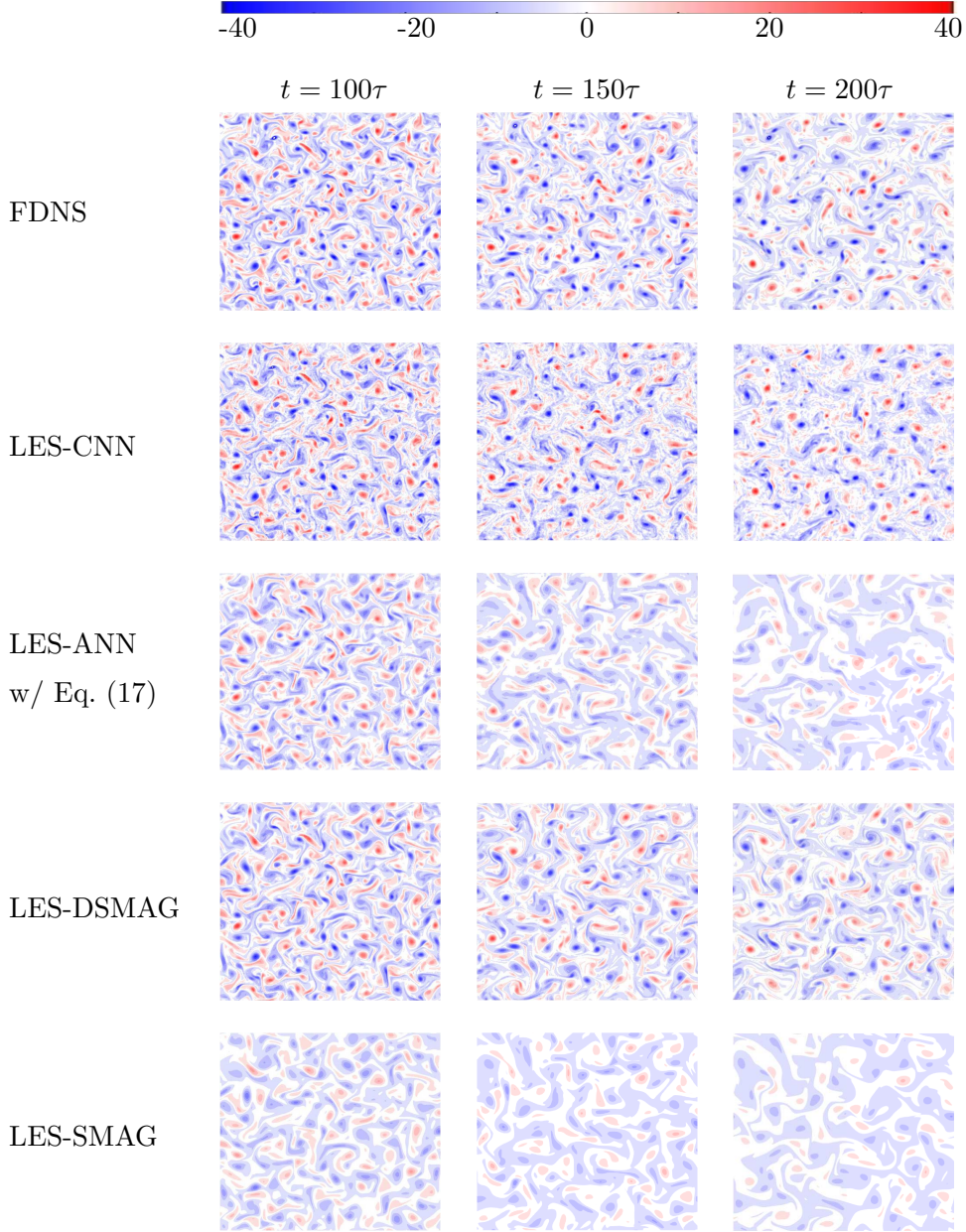
\begin{figure}[t]
\vspace{.1in}
 \centering
 \begin{overpic}[width=0.6\linewidth,height=1\linewidth]{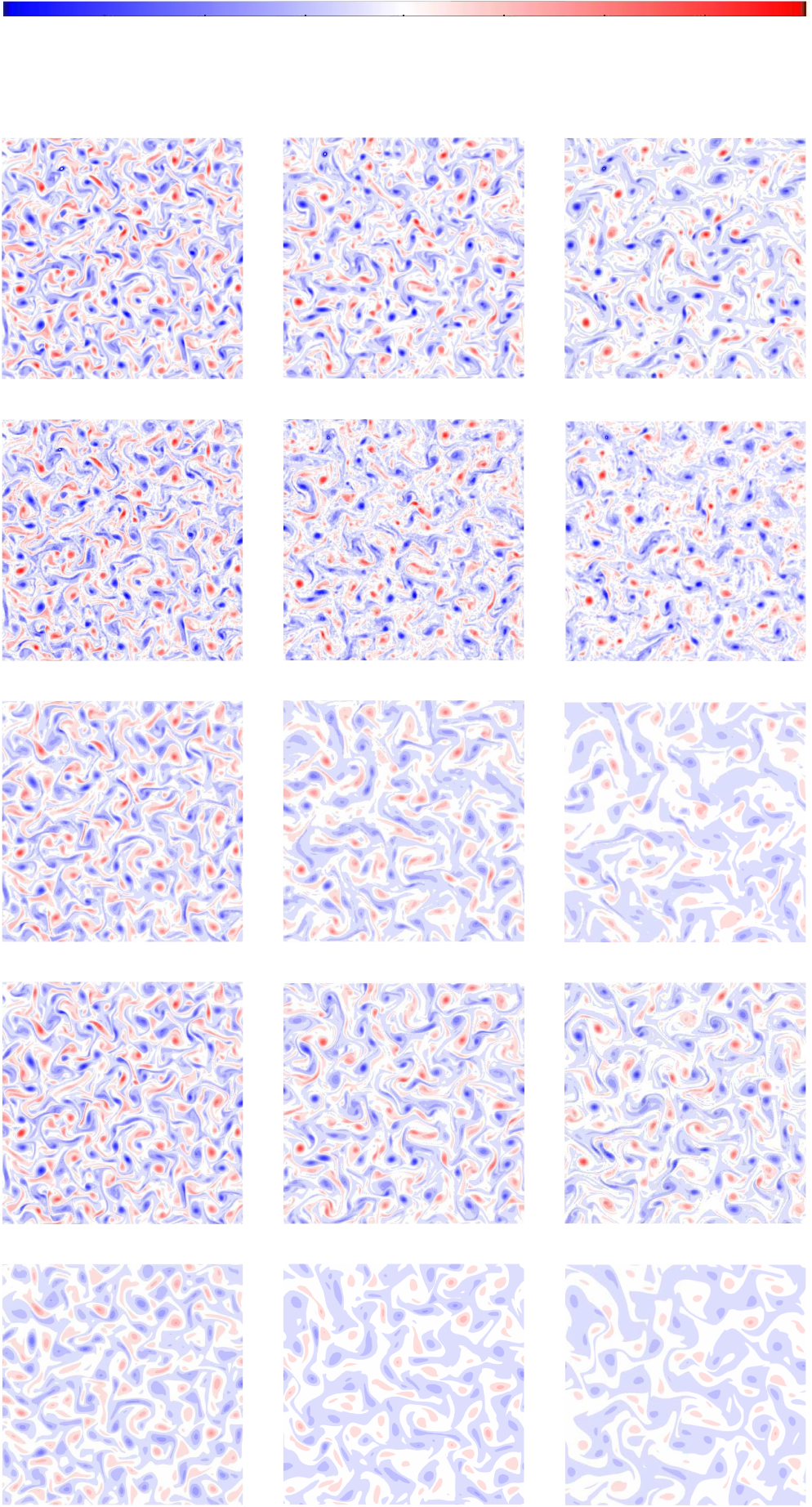}
 \put(-17,82){FDNS}
 \put(-17,63){LES-CNN}
 \put(-17,47){LES-ANN}
 \put(-17,43) {w/ Eq.~\eqref{clip}}
 \put(-17,25.5){LES-DSMAG}
 \put(-17,7){LES-SMAG}

 \put(0,97){-40}
 \put(14.5,97){-20}
 \put(29.3,97){0}
 \put(43.5,97){20}
 \put(58,97){40}

 \put(5,92){$t=100\tau$}
 \put(25.5,92){$t=150\tau$}
  \put(46,92){$t=200\tau$}

 \end{overpic}
 \caption{\small Examples of the vorticity fields at $t = 100\tau, 150\tau, \text{and }200\tau$ from one of the testing sets at $Re = 32000$. $\overline{\omega}$ from FDNS is shown in the first row (used as the ``truth'' for the LES). Rows 2-5 show $\overline{\omega}$ predicted from $t=50\tau$ using 4 \textit{a posteriori} LES models. The LES-CNN captures the patterns and magnitudes of both large- and small-scale structures well, except at the latest time at $t=200\tau$. While LES-DSMAG predicts most of the large-scale structures and some of the small-scale structures well, particularly at the earlier times, its overall accuracy is lower than that of LES-CNN (also see Fig.~\ref{fig:7}). The post-processed LES-ANN has a reasonably good performance at $t = 100\tau$, but at later times, this model and the LES-SMAG model are too diffusive such that the magnitude of the vorticity field is underpredicted and small-scale structures are missing.}
 \label{fig:6}
\end{figure}

To examine the accuracy of LES-CNN in short-term forecasting, Fig.~\ref{fig:7} presents the relative $L_2$-norm error in the prediction of $\overline{\omega}$ averaged from 5 random initial conditions in the testing set from $t=50\tau$ to $200\tau$. The results show that LES-CNN has the highest accuracy, outperforming the next best model, DSMAG. The post-processed LES-ANN and LES-SMAG have substantially higher errors, which as the next analysis shows is due to their excessive dissipation. To further evaluate the short-term accuracy of these LES models, Fig.~\ref{fig:6} shows an example of $\overline{\omega}(x,y)$ at $t=100\tau, 150\tau, \text{and } 200\tau$ predicted from an initial condition at $t=50\tau$ in the testing set. Evidently, LES-CNN is capable of predicting both small- and large-scale structures well, and outperforms LES-DSMAG, which while capturing most of the large-scale structures well, misses many of the small-scale structures. The post-processed LES-ANN and LES-SMAG are too diffusive and miss most small-scale structures, substantially underpredicting the magnitude of $\overline{\omega}$, especially at later times. \\

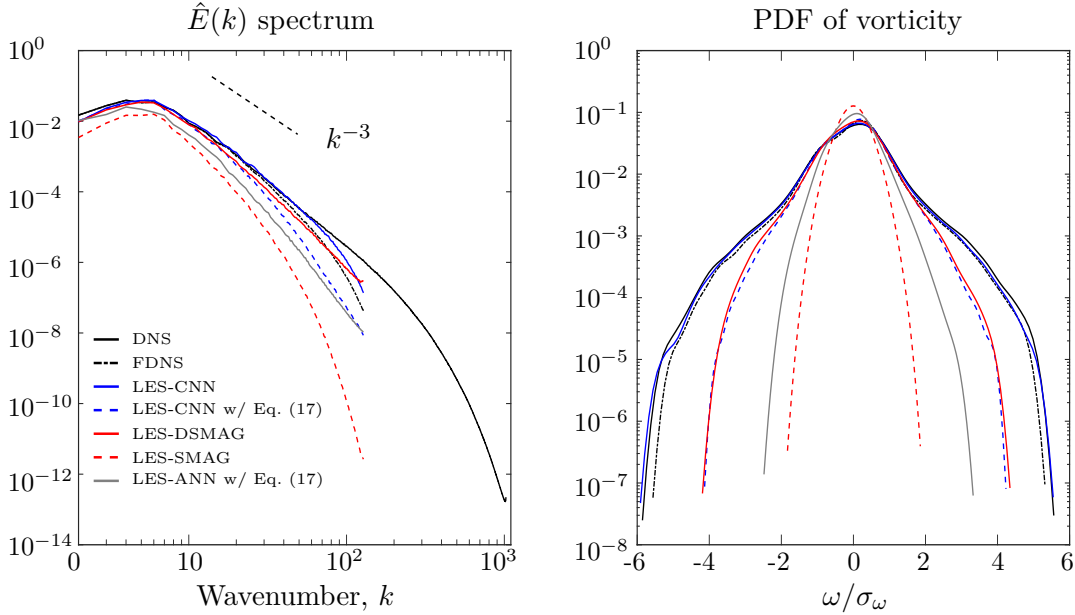
\begin{figure}[t]
\vspace{.1in}
 \centering
 \begin{overpic}[width=0.8\linewidth,height=0.4\linewidth]{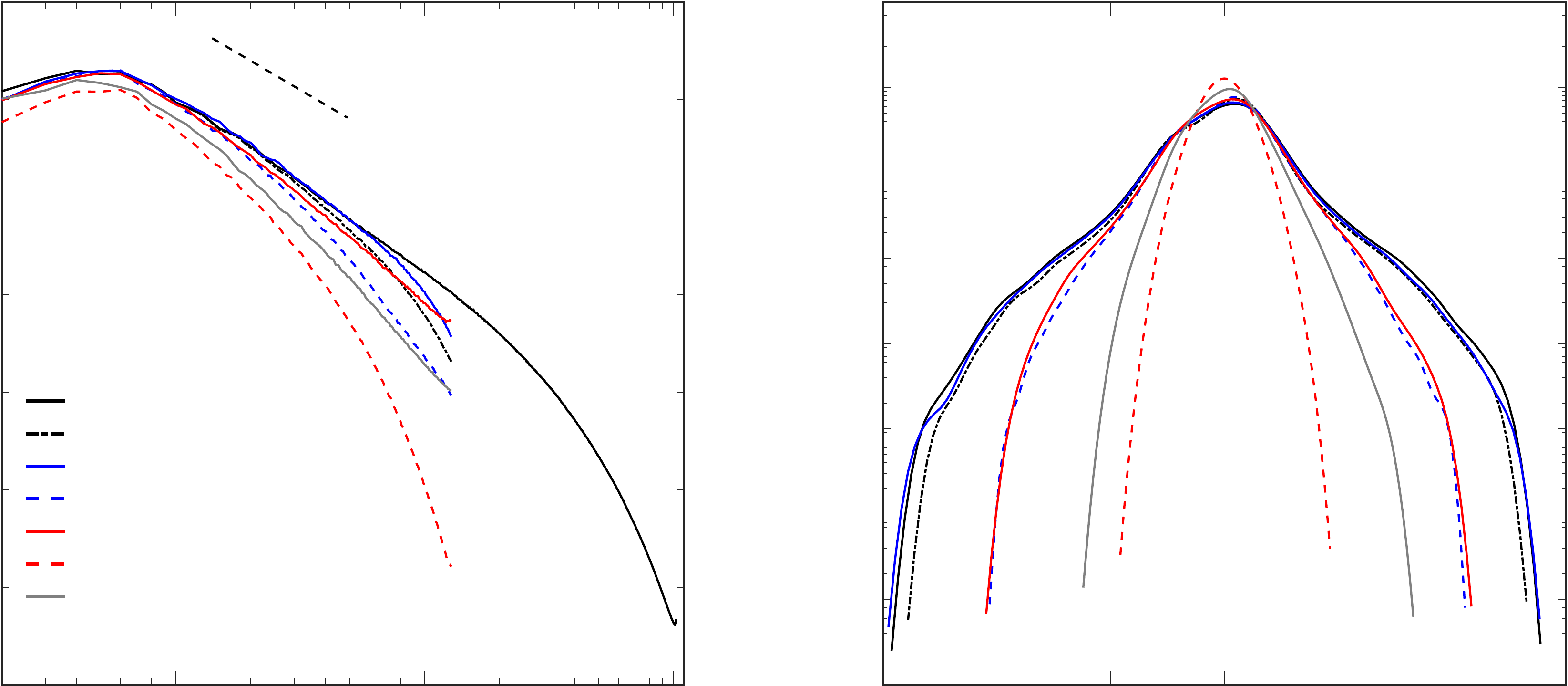}
 \put(68,52){PDF of vorticity}
 \put(11,52){$\hat{E}(k)$ spectrum}
 \put(25,40){$k^{-3}$}
 \put(75,-6){$\omega/\sigma_{\omega}$}
 \put(12,-6){Wavenumber, $k$}

 \put(50,-1.5){\small $10^{-8}$}
 \put(50,5){\small $10^{-7}$}
 \put(50,11.5){\small $10^{-6}$}
 \put(50,17.5){\small $10^{-5}$}
 \put(50,24){\small $10^{-4}$}
 \put(50,30){\small $10^{-3}$}
 \put(50,36){\small $10^{-2}$}
 \put(50,42.5){\small $10^{-1}$}
 \put(50,49){\small $10^{0}$}

 \put(55,-2.5){\small -6}
 \put(62,-2.5){\small -4}
 \put(69.5,-2.5){\small -2}
 \put(77.5,-2.5){\small 0}
 \put(84.5,-2.5){\small 2}
 \put(92,-2.5){\small 4}
 \put(99,-2.5){\small 6}

 \put(-7,-1){\small $10^{-14}$}
 \put(-7,5.5){\small $10^{-12}$}
 \put(-7,13){\small $10^{-10}$}
 \put(-7,20){\small $10^{-8}$}
 \put(-7,27){\small $10^{-6}$}
 \put(-7,34){\small $10^{-4}$}
 \put(-7,41){\small $10^{-2}$}
 \put(-7,49){\small $10^{0}$}

 \put(-1,-2.5){\small 0}
 \put(10,-2.5){\small 10}
 \put(25,-2.5){\small $10^2$}
 \put(40.5,-2.5){\small $10^3$}


 \put(5.5,20.3){\tiny DNS}
 \put(5.5,17.9){\tiny FDNS}
 \put(5.5,15.5){\tiny LES-CNN}
 \put(5.5,13.2){\tiny LES-CNN w/ Eq.~\eqref{clip}}
 \put(5.5,10.7){\tiny LES-DSMAG}
 \put(5.5,8.3){\tiny LES-SMAG}
 \put(5.5,6.2){\tiny LES-ANN w/ Eq.~\eqref{clip}}

 \end{overpic}
 \vspace*{10mm}
 \caption{\small The TKE spectrum $\hat{E}(k)$ and probability density function (PDF) of vorticity at $t = 200\tau$ from \textit{a posteriori} tests at $Re = 32000$. Results are from independent runs in the 5 testing sets. For $\hat{E}(k)$, the spectrum from each run is calculated and then averaged. For the PDF, data from all 5 runs are combined and the PDF is calculated using a kernel estimator~\cite{wilcox2010fundamentals}. For both the TKE spectrum and PDF, the LES-CNN has the best performance, followed by LES-DSMAG. Results from post-processed LES-CNN with Eq.~\eqref{clip} are shown just to demonstrate the importance of capturing backscattering for the excellent performance of LES-CNN in reproducing the TKE spectrum of FDNS and the tails of the FDNS' PDF. The post-processed LES-ANN with Eq.~\eqref{clip} and LES-SMAG are too diffusive, which shows in both TKE spectrum and PDF.}
 \label{fig:8}
\end{figure}

The above analysis shows that the superior accuracy of the CNN-based SGS model in \textit{a priori} tests translates to high accuracy in short-term forecasts with LES-CNN in \textit{a posteriori} tests. Next, we examine the accuracy of these \textit{a posteriori} LES models in reproducing the statistics of the turbulent flow, which is an important test for the applicability of these models~\cite{moser2020statistical}. Figure~\ref{fig:8} shows the TKE spectrum and probability density function (PDF) of vorticity at $t=200\tau$ from the 5 simulations in the testing sets. Among the LES models, LES-CNN has the best performance: its TKE spectrum matches that of the FDNS across wavenumbers and its PDF matches that of the FDNS, even at the end of the tails. The next best-performing model is LES-DSMAG, whose TKE spectrum overall agrees with FDNS, although this model is more diffusive than LES-CNN. The excessive diffusion is more noticeable in the PDF of the vorticity field: while the PDF of LES-DSMAG matches the bulk of the FDNS' PDF, there are large deviations at the tails, beyond $\pm 2$ standard deviations. The post-processed LES-ANN with Eq.~\eqref{clip} and LES-SMAG are too diffusive, leading to TKE spectra that quickly curl down as $k$ increases and PDFs that substantially deviate from the FDNS' PDF at the tails (for LES-SMAG, even in the bulk). Just to further demonstrate the importance of capturing backscattering in the outstanding performance of LES-CNN in matching the FDNS' spectrum and PDF, Fig.~\ref{fig:8} also presents results from a post-processed LES-CNN with Eq.~\eqref{clip} (i.e., backscattering removed), showing that the model becomes excessively diffusive (with performance comparable to that of the LES-DSMAG).\\

The \textit{a posteriori} results show the advantages of the CNN-based data-driven SGS model, which provides a stable LES model while capturing backscattering, and yields superior performance for both forecasting short-term spatio-temporal evolution and reproducing long-term statistics of the turbulent flow.


\subsection{Transfer learning to higher $Re$} \label{sec:TL}
\begin{figure}[t]
\vspace{.1in}
 \centering
 \begin{overpic}[width=0.8\linewidth,height=0.3\linewidth]{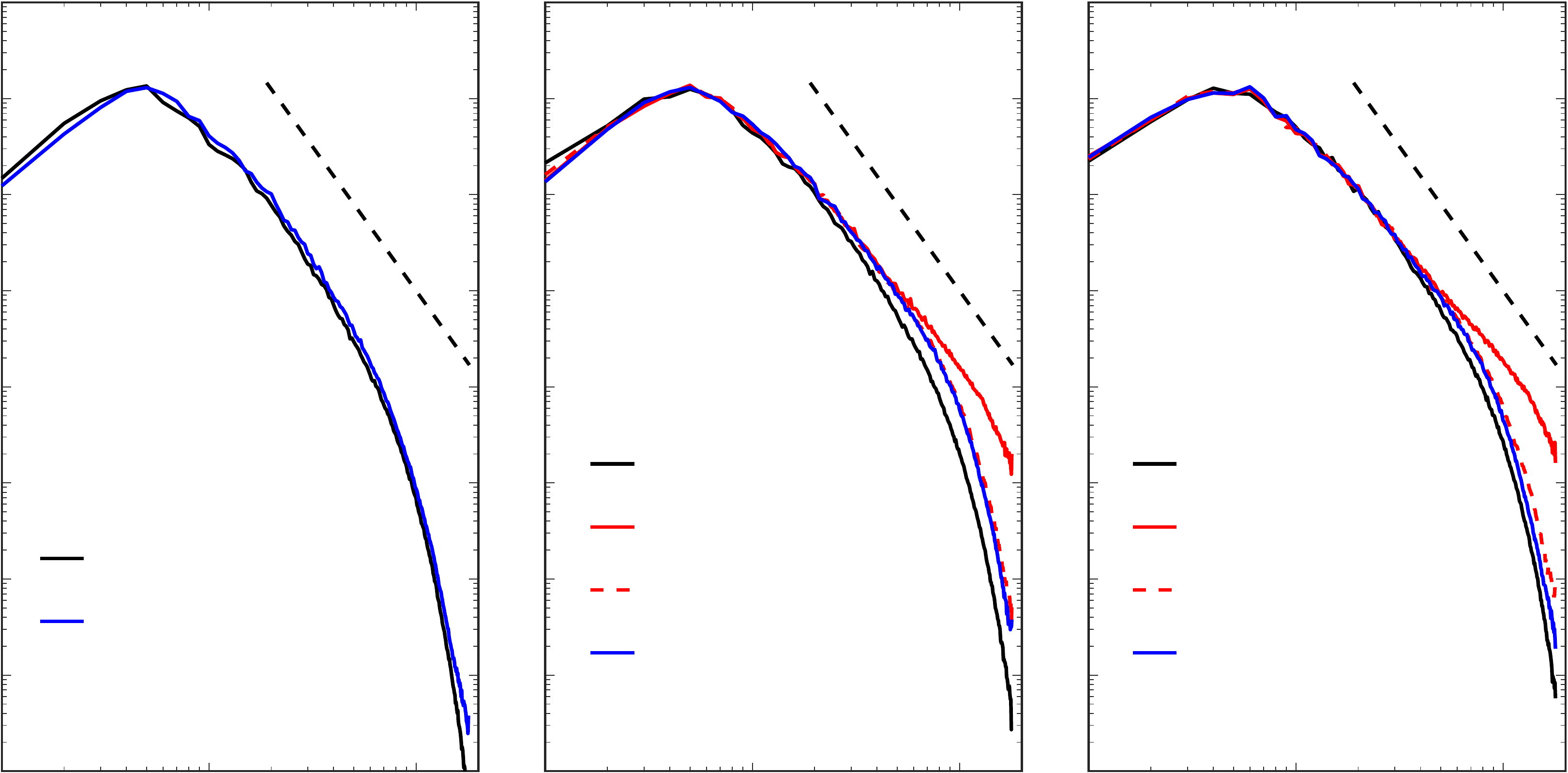}
 \put(-12,17){$\hat{E}(k)$}
 \put(3,39){Test on $Re=8000$}
 \put(37.5,39){Test on $Re=32000$}
 \put(72,39){Test on $Re=64000$}
 \put(93,30){$k^{-3}$}
 \put(57,30){$k^{-3}$}
 \put(22,30){$k^{-3}$}

 \put(6,10){\tiny FDNS}
 \put(6,6.9){\tiny LES-CNN$^{Re=8000}$}
 \put(41.5,14.5){\tiny FDNS}
 \put(41.5,11.5){\tiny LES-CNN$^{Re=8000}$}
 \put(41.5,8.3){\tiny LES-TL-CNN$^{Re=8000}$}
 \put(41.5,5.3){\tiny LES-CNN$^{Re=32000}$}

 \put(76.5,14.5){\tiny FDNS}
 \put(76.5,11.5){\tiny LES-CNN$^{Re=8000}$}
 \put(76.5,8.3){\tiny LES-TL-CNN$^{Re=8000}$}
 \put(76.5,5.3){\tiny LES-CNN$^{Re=64000}$}

 \put(-5,0){\tiny $10^{-8}$}
 \put(-5,8.5){\tiny $10^{-6}$}
 \put(-5,18){\tiny $10^{-4}$}
 \put(-5,27.5){\tiny $10^{-2}$}
 \put(-4,36.5){\tiny $10^{0}$}

 \put(-0.5,-2){\tiny $0$}
 \put(12,-2){\tiny $10^{1}$}
 \put(25,-2){\tiny $10^{2}$}

 \put(34,-2){\tiny $0$}
 \put(47,-2){\tiny $10^{1}$}
 \put(60,-2){\tiny $10^{2}$}

 \put(69,-2){\tiny $0$}
 \put(81,-2){\tiny $10^{1}$}
 \put(95,-2){\tiny $10^{2}$}

 \put(41,-5){Wavenumber, $k$}
 \end{overpic}
 \vspace*{7mm}
 \caption{\small Transfer learning to higher $Re$. The TKE spectrum $\hat{E}(k)$ at $t = 200\tau$ from \textit{a posteriori} tests at three different $Re$. Results are from independent runs in the 5 testing sets. For $\hat{E}(k)$, the spectrum from each run is calculated and then averaged. The superscript indicates the $Re$ on which the CNN is trained with $n_{tr}=50000$ samples. TL (transfer learned) means that the CNN has been re-trained with $n^{TL}_{tr}=500$ samples  ($1\%$ of $n_{tr}$) from the $Re$ on which the LES-CNN is tested on (indicated in the title of each panel). In each panel, the blue lines show that the LES-CNN trained and tested on the same $Re$ is accurate and its TKE spectrum agrees with that of the FDNS. However, the red lines in the two panels on the right show that the LES-CNN trained on $Re=8000$ does not perform well at $4\times$ or $8\times$ higher $Re$, with the TKE spectra of the simulated flow substantially deviating from that of the FDNS at high $k$ near $k_c$. The red dashed lines show that the LES-TL-CNN pre-trained on $Re=8000$ and transfer learned with a small amount of data from the higher $Re$ perform well at $4\times$ or $8\times$ higher $Re$.}
 \label{fig:11}
\end{figure}
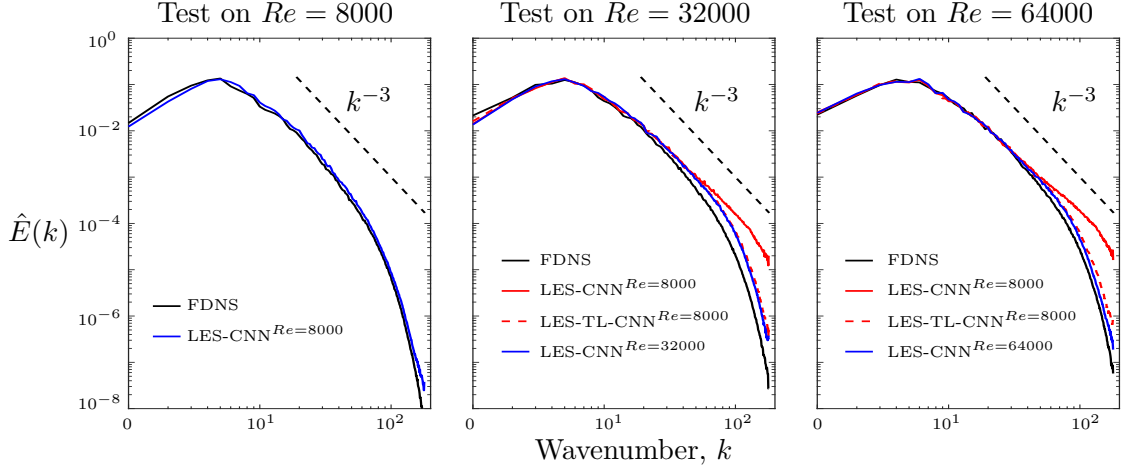

So far, we have tested the data-driven SGS model and the LES-CNN on flows with the same $Re$ as the flow from which data was collected for the training of the CNN ($Re = 8000$ or $Re = 32000$). As discussed in the Introduction, the capability to generalize beyond the training flow, in particular to extrapolate to turbulent flows with higher $Re$ in \textit{a posteriori} tests, is essential for robust, trustworthy, and practically useful LES models. Neural networks are known to have difficulty with extrapolations, and  in our recent work with multi-scale Lorenz 96 equations and forced 1D Burgers' turbulence, we found that data-driven SGS models do not generalize well to more chaotic systems or flows with $10\times$ higher $Re$, leading to inaccurate predictions in \textit{a posteriori} (online) tests~\cite{chattopadhyay2020data,subel2020data}. Similarly, Fig.~\ref{fig:11} shows that for the 2D-DHIT system studied here, a data-driven SGS model trained on data from $Re=8000$ leads to \textit{a posteriori} LES-CNN that is accurate only at $Re=8000$ but not at $Re=32000$ or $64000$. At these higher $Re$, the TKE spectra deviate substantially from the spectrum of the FDNS.  \\

In both Chattopadhyay~\textit{et al.}~\cite{chattopadhyay2020data} and Subel~\textit{et al.}~\cite{subel2020data}, we showed that transfer learning enables accurate generalization/extrapolation of data-driven SGS models to more chaotic systems and turbulent flows with a $10\times$ higher $Re$, although the effectiveness of this approach beyond 1D and to more complex turbulent flows remained to be investigated. \\

\begin{figure}[t]
\vspace{.1in}
 \centering
 \begin{overpic}[width=0.95\linewidth,height=0.12\linewidth]{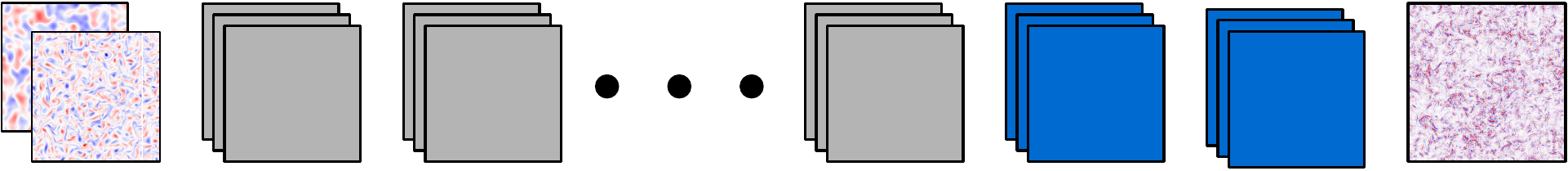}
 \put(-2,0){$\bar{\psi}$}
 \put(1,-2){$\bar{\omega}$}
 \put(94,-2){$\Pi$}

 \put(16,6){\textcolor{white}{Conv}}
 \put(16,3){\textcolor{white}{layer}}
  \put(29,6){\textcolor{white}{Conv}}
 \put(29,3){\textcolor{white}{layer}}
  \put(54.5,6){\textcolor{white}{Conv}}
 \put(54.5,3){\textcolor{white}{layer}}
  \put(67,6){\textcolor{white}{Conv}}
 \put(67,3){\textcolor{white}{layer}}
 \put(80,6){\textcolor{white}{Conv}}
 \put(80,3){\textcolor{white}{layer}}

 \end{overpic}
  \vspace*{5mm}
  \caption{\small Schematic of the CNN with transfer learning for extrapolation to higher $Re$. Everything is the same as the original CNN shown in Fig.~\ref{fig:3} with one exception: here, the first 8 Conv layers (gray) use the weights already computed during training with $n_{tr}$ samples from the lower $Re$ and are fixed (not to be trained). Only the last two Conv layers (blue) are going to be trained using $n^{TL}_{tr}=n_{tr}/100$ samples from the higher $Re$, after these layers are initialized not randomly but using the weights computed for the lower $Re$.}
 \label{fig:10}
\end{figure}
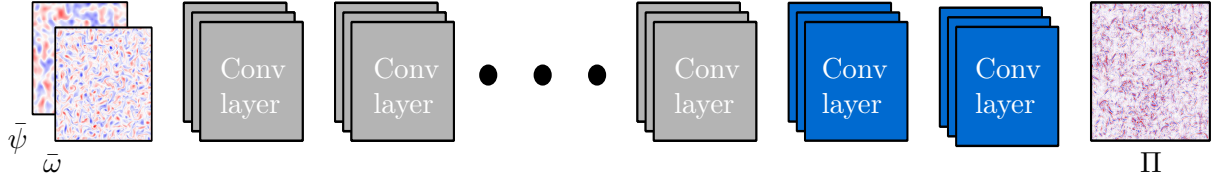

Transfer learning involves taking a neural network that has been already trained for a given data distribution (e.g., flow with a given $Re$) using a large amount of data and re-training only some of its layers (usually the deeper layers) using a small amount of data from the new data distribution (e.g., flow with a higher $Re$)~\cite{yosinski2014transferable,goodfellow2016deep}. For example, Fig.~\ref{fig:10} shows the schematic of the transfer-learned CNN used here. While similar to the original CNN (Fig.~\ref{fig:3}), there is one major difference: for transfer learning, the first 8 Conv layers use the weights already computed during training with $n_{tr}$ samples from the lower $Re$. These weights are fixed and remain the same during the re-training. The last two Conv layers are initialized with weights computed during training with $n_{tr}$ samples from the lower $Re$, but these two layers will be trained and their weights will be updated using $n^{TL}_{tr}=n_{tr}/100$ samples from the higher $Re$. The key idea of TL is that in deep neural networks, the first layers learn high-level features, and the low-level features that are specific to a particular data distribution are learned only in the deeper layers~\cite{yosinski2014transferable,goodfellow2016deep}. Therefore, for generalization, only the deeper layers need to be re-trained, which can be done using a small amount of data from the new distribution.  \\

To examine the effectiveness of transfer learning in the 2D-DHIT testcase, we take the CNN that is already trained with $n_{tr}$ samples from $Re=8000$ and re-train it with $n^{TL}_{tr}=n_{tr}/100$ samples from the flow with $Re=32000$ or $Re=64000$. Figure~\ref{fig:11} shows that the \textit{a posteriori} LES with these transfer-learned CNNs (LES-TL-CNN$^{Re=8000}$) accurately extrapolates to $4\times$ and $8\times Re$. In both cases, the accuracy of the transfer-learned LES-TL-CNN is as good as that of the LES-CNN trained with $n_{tr}$ samples from $Re=32000$ and $Re=64000$. Before showing the results for accurate extrapolation to even higher $Re$ ($16\times$) in the next section, we point out that the number of layers to be re-trained and the number of samples used for re-training  ($n^{TL}_{tr}$) depend on the problem and require some trial and error for the best performance. Here, fixing the first 6 layers and re-training the deeper 4 layers (with the same $n^{TL}_{tr}$) leads to similar LES-TL-CNN performance. The goal of transfer learning is to minimize $n^{TL}_{tr}$ while achieving the accuracy of $n_{tr}$, with the number of re-trained layers being a hyper-parameter to be tuned to achieve this goal. Substantial exploration in forced 1D Burgers' turbulence showed that the \textit{a posteriori} performance of LES with transfer-learned data-driven SGS models mainly depends on $n^{TL}_{tr}$ as long as more than one layer is re-trained~\cite{subel2020data}.

\subsection{Transfer learning to higher $Re$ and higher LES numerical resolution} \label{sec:TLRES}
One often-cited disadvantage of using CNNs (compared to local ANNs) for data-driven SGS modeling is dependence on the specific LES resolution for which the CNN has been trained, limiting the use of the LES-CNN on a different grid resolution (and the use of transfer learning to extrapolate to even higher $Re$ for which a higher LES resolution might be needed). Here, we show that this issue can be easily addressed by adding pooling (encoder) and upsampling (decoder) layers to the transfer learning architecture. \\

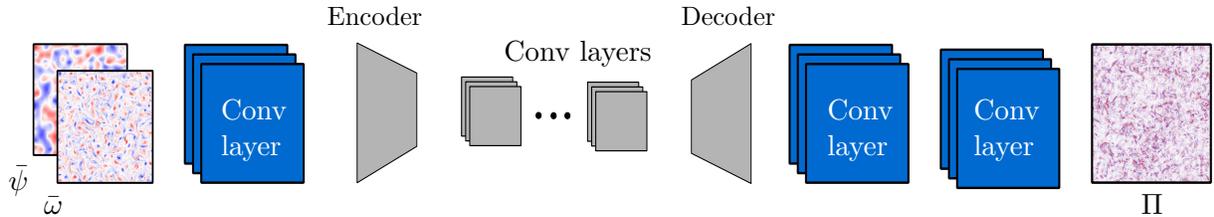
\begin{figure}[t]
\vspace{.1in}
 \centering
 \begin{overpic}[width=0.95\linewidth,height=0.12\linewidth]{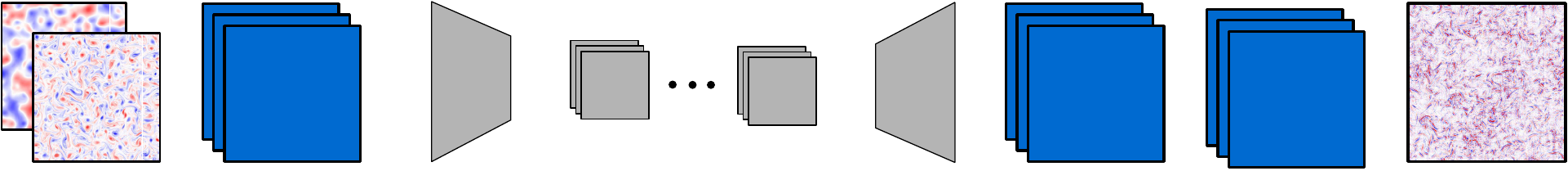}
 \put(-2,0){$\bar{\psi}$}
 \put(1,-2){$\bar{\omega}$}
 \put(94,-2){$\Pi$}

 \put(16,6){\textcolor{white}{Conv}}
 \put(16,3){\textcolor{white}{layer}}
 \put(25,14){\textcolor{black}{\small Encoder}}
 \put(40,11){\textcolor{black}{Conv layers}}
 \put(55,14){\textcolor{black}{\small Decoder}}
 \put(67.4,6){\textcolor{white}{Conv}}
 \put(67.4,3){\textcolor{white}{layer}}
 \put(79.8,6){\textcolor{white}{Conv}}
 \put(79.8,3){\textcolor{white}{layer}}

 \end{overpic}
 \vspace*{5mm}
\caption{\small Schematic of the CNN with transfer learning and encoder-decoder architecture for extrapolation to higher $Re$ and higher LES grid resolution. There are few differences with the CNN shown in Fig.~\ref{fig:10}. Here, the input and output samples are at the higher resolution of $512^2$ (inputs and outputs of the CNNs in Figs.~\ref{fig:3} and \ref{fig:10} are at the resolution of $256^2$). The 8 Conv layers that are already trained with $n_{tr}$ samples from the lower $Re$ and FDNS at the resolution of $256 \times 256$ are embedded within an encoder-decoder architecture. These 8 layers (gray) are fixed (not to be trained). The last two layers (in blue) are initialized not randomly but using the weights computed for the lower $Re$ and lower resolution. A first layer (in blue) is added between the input and the encoder, and is initialized randomly. Only these three layers are going to be trained using $n^{TL}_{tr}=n_{tr}/100$ samples from the higher $Re$ and higher resolution $(512^2)$.}
 \label{fig:12}
\end{figure}

For example, to use a CNN-based data-driven SGS model trained on data from $Re=8000$ and resolution $256 \times 256$ and conduct \textit{a posteriori} LES-TL-CNN integrations at $Re = 64000$ or $Re = 128000$ with resolution $N_\text{LES}=512$, we can use the encoder-decoder architecture shown in Fig.~\ref{fig:12}. Here, the number of convolutional layers are the same as before plus an additional layer before the encoder. The encoder with a pooling layer with stride two transforms the first layer from the input size ($512\times 512$) to the size of the layers of the CNN previously trained for a lower $Re$ and resolution ($256\times 256$). The 8 layers within the encoder-decoder have the weights already computed during training with $n_{tr}$ samples from the lower $Re$. These weights are kept fixed and these layers are not going to be trained. A decoder transforms the output of the last of these layers from the size $256\times 256$ to the size of the first of the last two layers, which is $512\times512$. Similar to Fig.~\ref{fig:10}, these two final layers are initialized with weights computed during training with $n_{tr}$ samples from the lower $Re$ (and lower resolution). Only these two layers and the very first layer will be trained and their weights are updated using $n^{TL}_{tr}=n_{tr}/100$ samples from the higher $Re$ and higher resolution. Here we use a factor of two increase in the resolution in each direction just as an example, and this approach can be used on any other resolution changes too.      \\


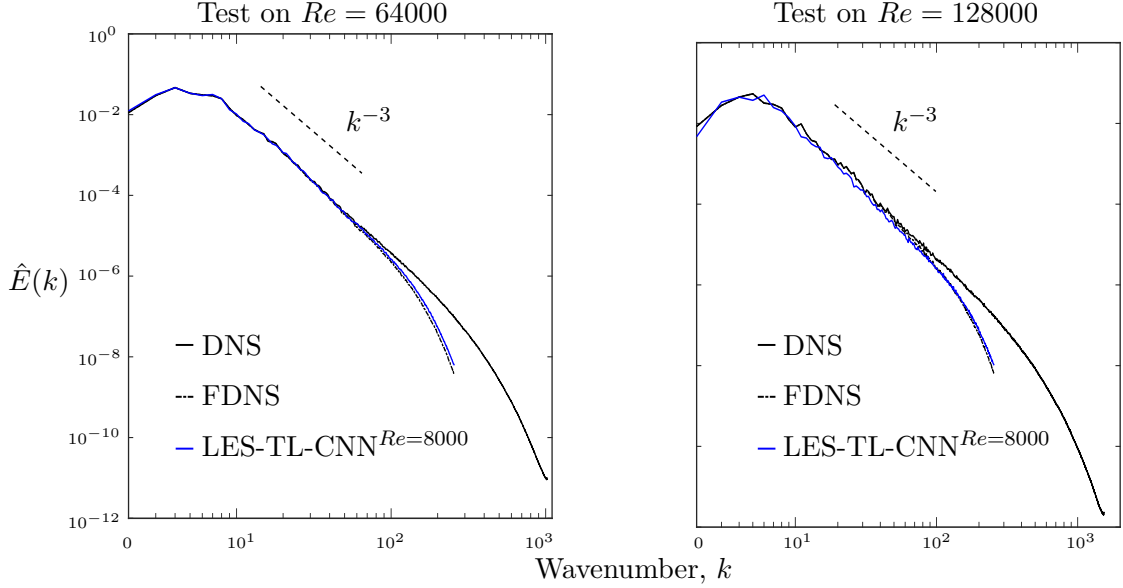
\begin{figure}[t]
\vspace{.1in}
 \centering
 \begin{overpic}[width=0.8\linewidth,height=0.4\linewidth]{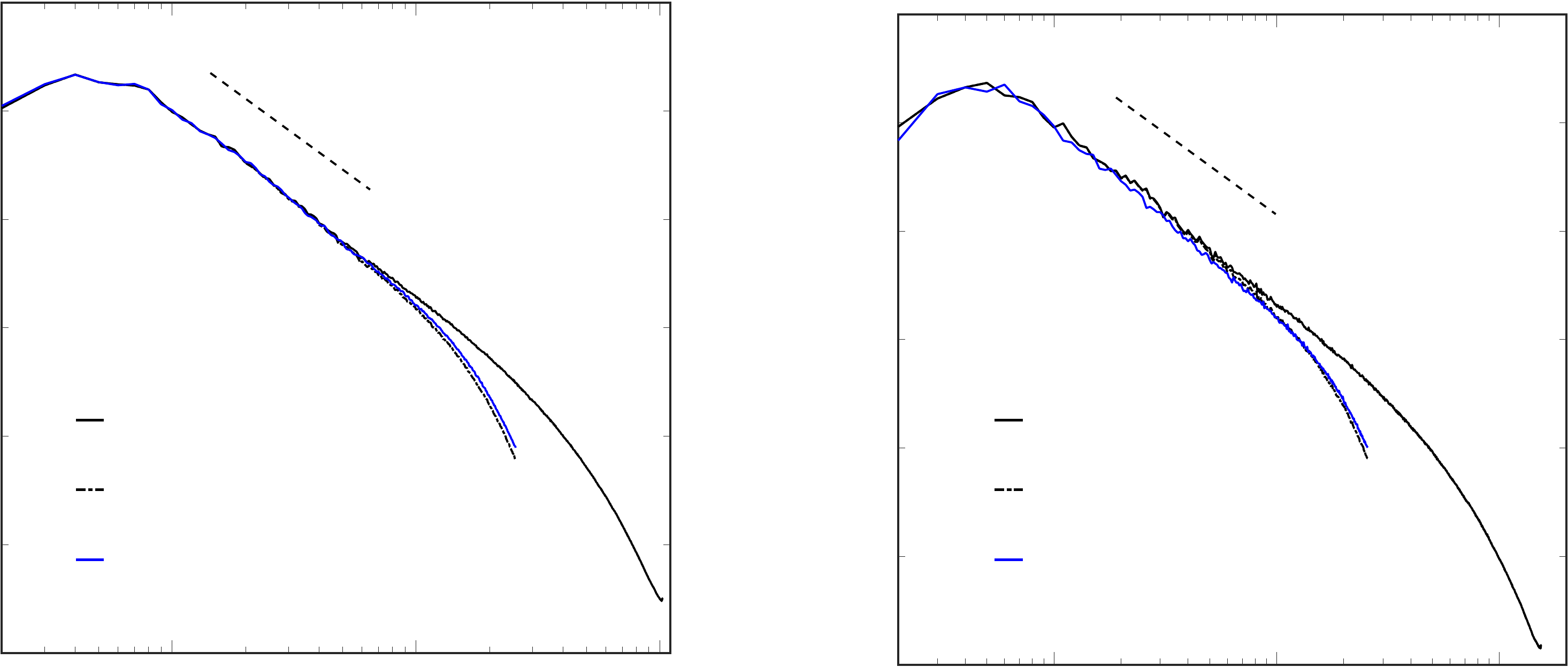}
 \put(7,51){Test on $Re = 64000$}
 \put(65,51){Test on $Re = 128000$}
 \put(22,40){$k^{-3}$}
 \put(77,40){$k^{-3}$}

 \put(7.5,17.5){DNS}
 \put(7.5,12.4){FDNS}
 \put(7.5,7.3){LES-TL-CNN$^{Re=8000}$}

 \put(66,17.5){DNS}
 \put(66,12.4){FDNS}
 \put(66,7.3){LES-TL-CNN$^{Re=8000}$}

 \put(-6,0){\tiny $10^{-12}$}
 \put(-6,8.5){\tiny $10^{-10}$}
 \put(-5,16.5){\tiny $10^{-8}$}
 \put(-5,25){\tiny $10^{-6}$}
 \put(-5,33){\tiny $10^{-4}$}
 \put(-5,41){\tiny $10^{-2}$}
 \put(-4,49){\tiny $10^{0}$}

 \put(-0.5,-2){\tiny $0$}
 \put(10,-2){\tiny $10^{1}$}
 \put(25,-2){\tiny $10^{2}$}
 \put(40,-2){\tiny $10^{3}$}

 \put(57,-2){\tiny $0$}
 \put(66,-2){\tiny $10^{1}$}
 \put(80,-2){\tiny $10^{2}$}
 \put(94.5,-2){\tiny $10^{3}$}

 \put(-12,24){$\hat{E}(k)$}
 \put(41,-5){Wavenumber, $k$}

 \end{overpic}
  \vspace*{10mm}
\caption{\small Transfer learning to higher $Re$ and higher LES numerical resolution. The TKE spectrum $\hat{E}(k)$ at $t = 200\tau$ from \textit{a posteriori} tests at two different $Re$. Results are from independent runs in the 5 testing sets. For $\hat{E}(k)$, the spectrum from each run is calculated and then averaged. The superscript indicates that the CNN has been trained with $n_{tr}=50000$ samples from $Re=8000$ at the resolution of $256\times256$. TL (transfer learned) means that the CNN has been re-trained with $n^{TL}_{tr}=500$ samples ($1\%$ of $n_{tr}$) from the $Re$ on which the LES-TL-CNN is tested on (indicated in the title of each panel) at the resolution of $512\times512$. In each panel, the spectra of the DNS and FDNS are shown; the latter is the ``truth'' for LES.  Note that for $Re = 64000$, $N_\text{DNS}=2048$ and for $Re = 128000$, $N_\text{DNS}=3072$. The FDNS is at the resolution of $512^2$. The blue lines show that the LES-TL-CNN pre-trained on $Re=8000$ and transfer learned with a small amount of data from the higher $Re$ and resolution perform well at $8\times$ or $16\times$ higher $Re$. Note that for LES-TL-CNN at both $Re$, here we use $N_\text{LES}=512$.}
 \label{fig:13}
\end{figure}

Figure~\ref{fig:13} shows, for $Re=64000$ and $Re=128000$, the TKE spectrum for LES-TL-CNN in comparison to that of FDNS. In this LES-TL-CNN, the numerical resolution is $N_\text{LES} \times N_\text{LES}=512\times 512$ and its CNN has been trained with $n_{tr}=50000$ samples from $Re=8000$ with resolution $256^2$ and transfer-learned with $n^{TL}_{tr}=n_{tr}/100$ samples from $Re=64000$ or $Re=128000$ at the resolution of $512^2$. The results show that transfer learning enables extrapolation to over an order-of-magnitude increase in $Re$ ($16\times$) and with the encoder-decoder architecture, enables transfer between different LES resolutions. The implications of these findings, in particular for practical purposes, are discussed in the next section.


\section{Summary and future directions} \label{sec:conclusions}
Using 2D decaying turbulence as the testbed, we have examined the performance of a CNN-based, non-local, data-driven SGS model in \textit{a priori} and \textit{a posteriori} analyses, with training and testing done on data from flows with the same $Re$. We have also investigated the effectiveness of transfer learning in enabling \textit{a posteriori} LES-CNNs that are trained on data from flows with low $Re$ (and low grid resolution) to work for flows with higher $Re$ (and higher grid resolution). In all cases, training is done on filtered DNS data, and the performance is tested in comparison with out-of-sample filtered DNS data.\\

As discussed in Section~\ref{a priori analysis}, \textit{a priori} tests at $Re=32000$ show that the trained data-driven SGS model can accurately predict the SGS forcing terms from never-seen-before snapshots of the resolved flow with correlation coefficients $c$ (Eq.~\eqref{cc}) around $0.93$, substantially outperforming a baseline physics-based SGS model, DSMAG. The data-driven SGS model is also found to accurately capture both forward transfer and backscattering between the resolved and unresolved scales.\\

To examine the connection between $\textit{a priori}$ and \textit{a posteriori} performance, we have evaluated the accuracy of \textit{a priori} tests (in terms of $c$) and the stability of \textit{a posteriori} LES-CNN as the number of training samples are varied from $n_{tr}=500$ to $50000$ (Table~\ref{table2}). This analysis shows that while the SGS model trained with $n_{tr}=10000$ seems accurate (with $c=0.90$), the LES with this CNN (and CNNs trained with smaller $n_{tr}$) is not stable. Increasing $n_{tr}$ to $30000$ and $50000$ further improves $c$ to $0.92$ and $0.93$, respectively, and leads to accurate and stable \textit{a posteriori} LES-CNN, without any need for post-processing or additional eddy viscosity. More analysis, in which $c$ is calculated separately for grid points experiencing only forward transfer or only backscattering, shows that at low $n_{tr}$, the CNN captures backscattering with much lower accuracy compared to forward transfer, but that the difference decreases as $n_{tr}$ is increased. This analysis suggests that the instabilities of \textit{a posteriori} LES-CNN trained with small training sets might be due to the inability of the SGS model to correctly represent backscattering. Why learning backscattering requires more data remains to be studied in future work. This might be because backscattering is fundamentally harder to learn data drivenly, or because backscattering is less frequent than forward transfer, or both. While we do not claim that all instabilities in \textit{a posteriori} (online) tests are due to this issue and could be overcome by increasing $n_{tr}$, we believe that these findings can help future studies in understanding the reasons(s) behind these instabilities and formulating rigorous solutions (see below for further discussions). \\

As discussed in Section~\ref{sec:post}, \textit{a posteriori} tests at $Re=32000$ with the CNN trained with $n_{tr}=50000$ show that LES-CNN is stable and accurate. The LES-CNN outperforms LES-DSMAG and LES with other tested SGS models in terms of both short-term forecast and re-producing the TKE spectrum and PDF of vorticity (even at the tails). The main shortcoming of the other models is that they are too diffusive, primarily because they do not capture backscattering due to their formulation or post-processing steps used to make them stable. The CNN-based SGS model learns both forward transfer and backscattering non-parameterically from data, and as mentioned above, once the latter is accurately captured with enough training samples, this SGS model leads to an accurate and stable $\textit{a posteriori}$ LES-CNN. \\

The analysis presented in Section~\ref{sec:TL} shows that a data-driven SGS model trained at $Re=8000$ does not lead to accurate \textit{a posteriori} LES-CNN solutions (in terms of TKE spectra) at the higher $Re$, e.g., at $Re=32000$ or $Re=64000$. However, we show that transfer learning largely solves this problem and enables the LES-CNN trained for a flow at low $Re$ to provide accurate and stable solutions for flows with higher $Re$ while requiring only a small amount of data from the flow at higher $Re$. The data-driven SGS model can even be coupled with LES solvers that use higher grid resolutions by adding an encoder-decoder architecture to the transfer-learned CNN (Section~\ref{sec:TLRES}). For example, we show that a CNN trained with $n_{tr}=50000$ samples from $Re=8000$ (at filtered resolution $256\times256$) can provide an accurate and stable \textit{a posteriori} LES-CNN for flows with $Re=128000$ and $N_\text{LES}=512$ once 2 out of the 10 convolution layers of the CNN are re-trained with only $n^{TL}_{tr}=n_{tr}/100=500$ samples from $Re=128000$. To the best of our knowledge, this is the first application of transfer learning to building generalizable data-driven SGS models beyond 1D turbulence (the 1D results were presented in our recent work~\cite{subel2020data}). \\

In summary, in a canonical 2D turbulent flow, we present promising results that CNNs and transfer learning can be used together to build non-local data-driven SGS models that lead to accurate, stable, and generalizable LES models. The generalization capability provided by transfer learning is key in making such data-driven SGS models practically useful. This is because training a base CNN model with a large training set of high-fidelity data from low $Re$ and then requiring only a small amount of high-fidelity data from the higher $Re$ for re-training is highly desirable for turbulence modeling, given the sharp increase in the computational cost of high-fidelity simulations such as DNS for higher $Re$. It should be also highlighted that because transfer learning only requires a small amount of data  and re-training only a few layers, its training process is fast and has a low computational cost, thus it can be conducted on the fly, for example when dealing with non-stationary systems. Moreover, the ability to also transfer between different LES resolutions further broadens the applicability of non-local SGS models. While not examined here, it is also possible that transfer learning provides generalization beyond $Re$ and grid resolution, for example between canonical fluid systems and fluid flows with more complex geometries. Such applications should be explored in future work.   \\

Beyond the obvious need to study the performance of the CNN-based SGS models and transfer learning in more complex turbulent flows (e.g., 3D, wall turbulence, stratified), there are a number of avenues to pursue in order to further expand and improve the methodology. The number of training samples might be potentially reduced, without loss of accuracy or stability, using data augmentation, e.g., through pre-processing the training data by exploiting the symmetries in the flow~\cite{pan2018long,subel2020data}, and/or using physics-informed ML~\cite{Kashinath2021}. Examples of the latter include adding components (such as capsules~\cite{chattopadhyay2020analog} and transformers~\cite{chattopadhyay2020deep}) that better preserve spatial correlations in the CNN or imposing physical constraints in the loss function~\cite[e.g.,][]{wu2020enforcing,Kashinath2021}. Establishing a connection between accuracy in \textit{a priori} tests and stability in \textit{a posteriori} tests would also be substantially helpful. Note that in this work (and in most other SGS modeling studies), an ``offline training'' strategy is used: the SGS model is first trained using snapshots of the resolved flow as inputs and snapshots of the SGS term as outputs, and then this trained data-driven model is coupled with the coarse-resolution LES solver. At least some of the issues related to stability could be potentially resolved, and even scaling with the size of training set could be improved, by using an ``online training'' strategy, which involves training the data-driven model to find the best SGS term that evolves the solution of the LES closest to that of the DNS. Sirignano~\textit{et al.}~\cite{sirignano2020dpm} have recently presented an exciting and promising framework for such an approach. Exploring data-driven SGS models that account for non-Markovian effects arising from coarse-graining, as suggested by the Mori-Zwanzig formalism~\cite{chorin2000optimal,wouters2013multi,li2015incorporation,parish2017dynamic}, is another direction to pursue in future work. Finally, interpreting the CNNs that provide accurate SGS models, such as the one trained here, can lead to insight into the SGS physics and possibly even better data-driven and/or physics-based models. While interpreting neural networks is currently challenging, using them along with data-driven equation discovery methods might provide a stepping stone, as for example done for ocean mesoscale eddies in pioneering work by Zanna and Bolton~\cite{zanna2020data}.

\section{Acknowledgments}
We thank Romit Maulik, Tapio Schneider, and Laure Zanna for insightful discussions about data-driven SGS modeling, and Karan Jakhar, Rambod Mojgani, and Ebrahim Nabizadeh for helpful comments on the manuscript. This work was supported by an award from the ONR Young Investigator Program (N00014-20-1-2722) and a grant from the NSF CSSI program (OAC-2005123) to P.H. Computational resources were provided by NSF XSEDE (allocation ATM170020) to use Bridges GPU and Comet GPU clusters and by the Rice University Center for Research Computing. The codes for CNN and CNN with transfer learning are publicly available at \url{https://github.com/envfluids/2D-DDP}.

\appendix
\section{Initial condition for DNS}\label{Initial conditions of DNS}

Following previous studies, we choose the initial conditions of DNS to have the same energy spectrum but randomly different vorticity fields~\cite{orlandi2012fluid,maulik2017stable,maulik2019subgrid}. The initial energy spectrum is given by \cite{orlandi2012fluid}
\begin{eqnarray}\label{Eini}
\hat{E}(k) = Ak^4e^{-(k/k_p)^2},
\end{eqnarray}
where the amplitude is
\begin{eqnarray}\label{Eini_A}
A = \frac{4k_p^{-5}}{3\pi},
\end{eqnarray}
and $k = |\mathbf{k}|=\sqrt{k_x^2+k_y^2}$. The maximum value of the energy spectrum occurs at $\sqrt{2} k_p$, where $k_p=10$ is used here following Ref.~\cite{maulik2017stable}. The given energy spectrum in turn determines the magnitude of the Fourier coefficients of vorticity:
\begin{eqnarray}\label{Vor_Mag}
|\hat{\omega}(\mathbf{k})| = \sqrt{\frac{k}{\pi}\hat{E}(k)}.
\end{eqnarray}
Then the vorticity distribution in Fourier space is
\begin{eqnarray}\label{Vor_Mag}
\hat{\omega}(\mathbf{k})=|\hat{\omega}(\mathbf{k})|e^{\mathbf{i}\eta(\mathbf{k})},
\end{eqnarray}
where $\eta(\mathbf{k}) = \eta_1(\mathbf{k}) + \eta_2(\mathbf{k})$. $\eta_1(\mathbf{k})$ and $\eta_2(\mathbf{k})$ are independent random numbers from a uniform distribution in $[0,2\pi]$ at each $(k_x,k_y)$ when both $k_x, k_y\geq 0$ (first quadrant of the $k_x-k_y$ plane). The values at the other quadrants are as follows:
\begin{subequations}\label{initial-vorticity}
\begin{eqnarray}
\eta(\mathbf{k}) &=& -\eta_1(\mathbf{k}) + \eta_2(\mathbf{k}) \text{\;for\;} k_x<0, k_y\geq0\label{IV2}\\
\eta(\mathbf{k}) &=& -\eta_1(\mathbf{k}) - \eta_2(\mathbf{k}) \text{\;for\;} k_x<0, k_y<0\label{IV3}\\
\eta(\mathbf{k}) &=& +\eta_1(\mathbf{k}) - \eta_2(\mathbf{k}) \text{\;for\;} k_x\geq0, k_y<0\label{IV4}
\end{eqnarray}
\end{subequations}
The initial vorticity field is applied at $t=0$.

Figures~\ref{fig:1}(a) and \ref{fig:1}(d) show an example of the initial $\omega(x,y)$ and the corresponding $\hat{E}(k)$, respectively. The initial vorticity is dominated by relatively large-scale structures, but small-scale structures emerge as the flow evolves  (Figs.~\ref{fig:1}(b), (c), and (d)). From $t \approx 50\tau$, the $\hat{E}(k)$ spectrum exhibits self-similarity and follows the Kraichnan-Batchelor-Leith (KBL) theory~\cite{kraichnan1967inertial,batchelor1969computation,leith1971atmospheric}. Between $t = 50\tau$ and $200\tau$, the flow decays due to the viscous dissipation, the small-scale structures fade away, and the large, coherent vortices merge and grow as a result of the inverse energy cascade. Following previous studies, we focus on this phase of the decaying 2D turbulence and discard the first $50\tau$ as spin-up~\cite{maulik2017stable,beck2019deep}. 

\pagebreak
\setlength{\bibsep}{2.6pt plus 1ex}
\begin{spacing}{.01}
	\small
\bibliographystyle{abbrv}
\bibliography{main3}
\end{spacing}
\end{document}